\documentclass[
a4paper,
reprint,
showkeys,
nofootinbib,
twoside,
notitlepage
]{revtex4-1}

\usepackage[
centering,
hmargin=2cm,
vmargin=2cm
]{geometry}

\usepackage[T1]{fontenc}
\usepackage{graphicx}
\usepackage{soul}
\usepackage{subfigure}
\usepackage{xcolor}
\usepackage{booktabs}

\usepackage{
amsmath,
amssymb,
mathtools,
bm,
amsthm,
braket,
tensor,
mhchem,
nccmath
}

\usepackage{enumitem}

\usepackage[
colorlinks=true,
linkcolor=blue,
citecolor=blue,
urlcolor=blue
]{hyperref}

\usepackage{cleveref}

\begin{document}

\title{An Asymptotically Causal Metamodel for Neutron Star Equations of State}

\author{Gabriele Montefusco$^{a}$} 
\author{Marco Antonelli$^{a}$} 
\author{Francesca Gulminelli$^{a,b}$} 
\affiliation{
\vspace{1mm}
\makebox[0pt]{$^{a}$ Université de Caen Normandie, CNRS/IN2P3, LPC Caen UMR6534, 14050 Caen, FR}
\\
\makebox[0pt]{$^{b}$ Institut Universitaire de France (IUF), FR}
}

\begin{abstract}
Nuclear metamodels - phenomenological parametrizations of the energy of nuclear matter - are convenient tools to explore the space of realistic neutron star configurations constrained by astrophysical and nuclear data. While much recent work has focused on composition-agnostic barotropic models, the metamodel approach is designed to describe the composition dependence of the relevant thermodynamic potential. We revise a previously proposed non-relativistic metamodel by introducing a more controlled high-density behavior, improving both its causal properties and its accuracy in reproducing the pressure and the $\beta$-equilibrium composition of microscopically motivated equations of state. Since asymptotic causality is enforced by construction, the fraction of discarded models due to superluminal sound speeds is substantially reduced, facilitating metamodel-based explorations of equilibrium neutron star configurations. We further assess our framework by performing a Bayesian inference of neutron star properties beyond standard observables such as masses and radii, exploiting the metamodel's ability to probe composition-dependent quantities including the dUrca threshold and the Ledoux criterion for g-mode stability.
\end{abstract} 

\maketitle

\section{Introduction}
\label{intro}

Observational progress in the last decade has led to increasingly precise determinations of neutron stars' macroscopic properties, in particular masses and radii inferred from X-ray pulse-profile modeling \cite{Nicer1,Nicer2,Riley_2019,Riley_2021,Salmi2022,NicerJ0030,Rutherford_2024,NicerJ0437}.
Independent constraints were also provided by gravitational-wave observations of compact binary inspirals by the LIGO-Virgo-KAGRA (LVK) collaboration \cite{TheLIGOScientific_2017,GW170817_1,AbbottPRX2023}, with further improvements expected from next-generation detectors \cite{MaggioreJCAP2020,Evans2021,BranchesiJCAP2023,Abac2025}. 
Interpreting these data requires a reliable description of the equation of state (EoS) of dense matter and its uncertainties.

Apart from providing a link between the microphysics of dense matter and static neutron star (NS) observables, the EoS is also a fundamental ingredient in simulations of NS dynamics, from mergers \citep{baiotti_rezzolla_review_2017,radice_bernuzzi_perego_2020,ecker_2025} to cooling~\citep{Potekhin2015review,Marino_2024} and pulsar glitches~\citep{montoli2020A&A,AMP_arxiv_2023}.
This has motivated analyses that combine astrophysical observations with data from nuclear structure and heavy-ion collisions to constrain the NS EoS~\citep{AnnalaPRL2018,TewsEPJA2019,huth2022Natur,TsangNAT2024,MUSES_2024review,Koehn_2025}.

Many current EoS inferences rely on protocols that systematically explore the space of pressure-energy density relations of dense matter independently of its composition. Examples of such composition-agnostic approaches include piecewise polytropes \citep{Read_2008,suleiman_2022}, spectral representations~\citep{lindblom_2010spectral,Lindblom_2012}, pressure-density nodes \citep{ozel_psaltis_2009}, sound speed parametrizations~\citep{alford_constant_speed_2015,tews_2018apj,GreifMNRAS2019,oboyle_prd_2020,Brandes_sound_inference_23}, and nonparametric representations based on Gaussian processes~\citep{Landry_2018,Essick_2019}.
These composition-agnostic protocols are optimal EoS inference strategies when the data consist of global static NS observables (masses, tidal deformabilities, and radii \citep{sun_lattimer_2026arXiv}), as such observables depend only on an effectively barotropic EoS, that is, on the chemically equilibrated section of extended EoS models with additional compositional degrees of freedom; see \cite{Haensel2007book,oertel_RMP_17,burgiofantina2018,Burgio_eos_2021} for a review.

Accordingly, the most robust inferences of the NS EoS to date have been based on large ensembles of phenomenological EoSs that are composition-agnostic. However, while such EoS representations are adequate for static global NS observables, they cannot describe composition-dependent quantities such as the direct Urca threshold \citep{lattimerUrca1991,klahn_PhysRevC_2006}, convective stability via the Schwarzschild-Ledoux discriminant \citep{Lai1994}, or departures from $\beta$-equilibrium relevant for transport coefficients (in particular, bulk viscosity~\citep{Sawyer89,gavassino_bulk_2021,camelio_I,harris_Bulk_viscosity_2024arXiv,yang2025PhRvL,harris2025prc}) and their effect on quasi-normal oscillation modes~\citep{finn1987,Reisenegger_1992,Counsell2024MNRAS,Montefusco2025,zhao_haber_2025ApJ}.
Their use is also limited in NS regions pertaining to the solid crust and the crust-core transition, which can introduce systematic effects on global static properties~\cite{fortin_and_everyone_2016,Ferreira2020,Suleiman2021,davis2024,burrello2025crust,klausner2025prc}. 

To overcome these limitations of composition-agnostic schemes, one may assume a certain composition of matter (i.e., the number and nature of the matter fields) while remaining agnostic about the exact microphysics involved (i.e., the effective many-body Hamiltonian or Lagrangian). In practice, this means extending the agnostic approach from pressure-energy relations to energy-composition relations. This is the idea behind phenomenological metamodels, which are composition-aware parametrizations of the macroscopic energy of nuclear matter valid within a range of baryon density and isospin asymmetry relevant for both nuclear and NS applications~\citep{MargueronMetaI,MargueronMetaII,Huth2021,Lim2024}. 
Their utility is twofold. On the one hand, they provide an analytic and computationally efficient framework in which current nuclear and astrophysical information can be propagated to composition-dependent quantities, making explicit which quantities can be reasonably predicted and which ones remain weakly constrained by current data. On the other hand, they can serve as a null hypothesis for the possible appearance of exotic degrees of freedom or first-order phase transitions \citep{HoaUniverse}, which are expected on theoretical grounds at high density, even though their presence cannot yet be established from current astrophysical data \citep{Brandes_2023,Brandes_2025}. For hypothesis testing, an essential requirement for a nucleonic metamodel is a sufficiently large parameter space to accurately reproduce different families of microscopic or phenomenological nucleonic models and interpolate among them. 

In particular, the nucleonic metamodel of \citet{MargueronMetaI} has been widely used in studies of NS global properties \citep{Tews_2018,CarreauMeta,HoaUniverse,Somasundaram_2021,davis2024,Chatterjee_2026}, crust composition and the nature of the crust-core transition \citep{AnticJPG2019,HoaPasta,Grams_2022,grams_diverres_2025,klausner2025prc,burrello2025crust,Suleiman_2025,BaoAn_2025,BaoAn_2026,Klausner_prc_2026,diverres_shear_2026}, and NS oscillation mode analysis~\citep{Montefusco2025}.

However, the original metamodel parametrization of \citet{MargueronMetaI} is known to produce superluminal sound speeds or mechanical instabilities at high densities of a few times nuclear saturation, requiring strong parameter restrictions or explicit density cutoffs~\citep{HoaUniverse,Montefusco2025}. For this reason, it is sometimes used only up to a certain breaking density, above which more robust agnostic modeling is employed~\citep{Suleiman_2025,Koehn_2025}.
Such truncations hinder the exploration of high-density physics and NS structures under the null hypothesis of a unified nucleonic model, a possibility that is useful in hypothesis testing and prospective studies~\citep{mondal2023MNRAS}.

To overcome this, one may implement the metamodel philosophy (i.e., a composition-aware procedure with parameters that can be varied broadly to probe the EoS space) within relativistic mean-field (RMF) theory, as this framework is more likely to enforce causality by construction \citep[e.g.][]{li_sedrakian_2019,Malik_2022,li_sedrakian_2023,Malik_2024,MalikSurvey}. 
The downside is that RMF protocols are more computationally expensive for the large ensembles required in Bayesian inference. Additionally, the required model flexibility at high density encourages the use of density-dependent couplings with a complex density dependence~\citep{ScurtoPrediction,char_metaRMF_2023}, which again may lead to unphysical instabilities~\citep{scurto2025delta}.

This leaves a gap between the algebraic simplicity of the original metamodel scheme of \citet{MargueronMetaII} and the robustness of the RMF at high density.
In this work, we construct a nucleonic metamodel that fills this gap: unlike RMF-based metamodels, it has an analytic structure that allows for an exact mapping between some of its parameters and standard nuclear matter parameters (NMPs), while modifying the density dependence of the interaction terms to ensure causal and stable behavior up to central NS densities and to precisely reproduce realistic EoSs across Skyrme-like and RMF-like families, including composition-related observables. 

\section{Metamodel representation of the nucleonic EoS}

We recall some known facts about the nuclear EoS. 
This serves both to set the stage and to define our notation and premises.

\subsection{Preliminary definitions and metamodel ansatz}
\label{sec_prelimin}

The key quantity we aim to model is the energy density (or, equivalently, the energy per baryon) of homogeneous and isotropic nuclear matter in which only strong interactions are considered: baryon number and isospin are conserved and, in the absence of external fields or persistent currents \citep{thermo_2020CQG}, the zero-temperature energy density $\epsilon(n_n,n_p)$ depends only on the neutron and proton number densities.

A \emph{nucleonic metamodel} is a parametrization $\epsilon_X(n_n,n_p)$ of the (unknown) energy density such that suitable choices of the parameters $X$ exist for which $\epsilon_X(n_n,n_p)$ is consistent with selected information from nuclear experiments or theoretical studies, such as current estimates of nuclear matter parameters (NMPs) at saturation~\citep{rocamaza2018PrPNP,Vidana2019nqn,BaoAn_2021,Lattimer2023constraints}.

In practice, the parameters $X$ may be tuned to reproduce a given microscopically motivated $\epsilon(n_n,n_p)$, or varied to explore classes of admissible nucleonic EoSs beyond those present in the literature. 
Any reasonable metamodel scheme designed to operate over a given range of baryon density and isospin asymmetry should span the space of nucleonic EoSs, or provide an analytic fit to a tabulated $\epsilon(n_n,n_p)$ within its domain of validity. 
Hence, a metamodel is essentially a microscopically agnostic but composition-aware procedure for spanning the space of reasonable dense-matter EoSs.

Once $\epsilon_X(n_n,n_p)$ is specified, the corresponding chemical potentials and pressure follow from standard thermodynamic relations,
\begin{equation}
\label{eq_thermo}
\mu_X^q = \partial_q \epsilon_X,
\qquad
P_X = \sum_{q=n,p} n_q \mu_X^q - \epsilon_X ,
\end{equation}
where $q=n,p$ labels neutrons and protons and $\partial_q$ is the partial derivative with respect to $n_q$. Equivalently, introducing the total baryon density $n=n_n+n_p$ and the fractions $x_q=n_q/n$, the pressure also reads
\begin{equation}
\label{eq_PX}
P_X(n_n,n_p) = n^2 \frac{d}{dn} \frac{\epsilon_X(n x_n,n x_p)}{n} \,.
\end{equation}
A metamodel is defined once the functional form of $\epsilon_X$ (or any other convenient thermodynamic potential) is specified, and the other thermodynamic quantities are consistently derived via \eqref{eq_thermo}.
The original metamodel~\citep{MargueronMetaI} builds on a specific ansatz for $\epsilon_X$, which we generalize here as
\begin{align}
\label{eq_meta_ansatz}
\epsilon_X( n_n , n_p ) = \epsilon_F(n, \delta) + n \sum_i u_i(n) \, \delta^i \, ,
\end{align}
where $\delta = (n_n-n_p)/n = 1-2 x_p$ is the isospin asymmetry and $i$ runs over a subset of the non-negative integers.
In \eqref{eq_meta_ansatz}, $\epsilon_F(n, \delta) = \epsilon_F^n(n_n)+\epsilon_F^p(n_p)$ is the energy density of a non-interacting Fermi gas mixture, with
\begin{equation}
\label{eq_eps_F}
\begin{split}
\epsilon_F^q(n_q) = \frac{m_q^4}{8 \pi^2 \kappa^3 } 
\left[ y_q (2y_q^2 + 1) \sqrt{y_q^2 + 1}-
\right. \\ \left. 
- \ln \left( y_q + \sqrt{y_q^2 + 1} \right) \right]   ,
\end{split}
\end{equation}
where $\kappa=\hbar c = 197.32\,$MeV$\,$fm, $y_q=\kappa k_q/m_q$ is the relativity parameter, and $k_q = (3 \pi^2 n_q)^{1/3}$ is the Fermi wave vector of each species~\citep{Jancovici1962NCim,Faussurier2016}. 
Note that, although the original metamodel includes effective nucleon masses, in \eqref{eq_eps_F} and throughout this work we use only the physical masses $m_n =  939.57\,$MeV and $m_p = 938.28\,$MeV. An extension of \eqref{eq_meta_ansatz} to incorporate phenomenological effective masses into $\epsilon_F(n,\delta)$ would be important for finite-temperature applications, but is not required for the purposes of this study.

The terms $u_i(n)$ in \eqref{eq_meta_ansatz} are necessary to obtain a realistic energy per baryon of interacting nuclear matter at arbitrary asymmetries $\delta$ and densities~$n$: 
\begin{equation}
\label{eq_eX}
    e_X(n_n,n_p) = \frac{1}{n}\left[ \epsilon_X(n_n,n_p)-\sum_q m_q n_q   \right] .
\end{equation}
Note that the energy density \eqref{eq_meta_ansatz} contains mass terms via \eqref{eq_eps_F}, which are then subtracted in~\eqref{eq_eX}. 

Using \eqref{eq_PX}, the pressure is 
\begin{equation}
    \label{eq_PXeuler}
    P_X(n_n,n_p) = \sum_q P_F^q(n_q)+n^2\sum_i \frac{du_i}{dn}\delta^i \, ,
\end{equation}
where $P_F^q(n_q)$ are the Fermi-gas contributions, 
\begin{equation}
\label{eq_PF}
\begin{split}
    P_F^q(n_q) = \frac{m_q^4}{24 \pi^2 \kappa^3}\left[ 
    y_q (2y_q^2 - 3)\sqrt{y_q^2 +1}+ \right.
    \\
    \left. +3  \ln \left( y_q + \sqrt{y_q^2 + 1} \right) 
    \right]  \, .
    \end{split}
\end{equation}

\subsection{Physical and practical requirements}

The usefulness of a metamodel scheme hinges on its ability to satisfy a set of physical and practical requirements, such as controlled low-density limits~\citep{burrello2025crust}, consistency with empirical distributions of NMPs~\citep{xu2022PhRvC,Klausner_2024,Klausner_prc_2026}, and the enforcement of thermodynamic stability and causality over the density and composition ranges relevant for nuclear and NS applications~\citep{Lim2024,Montefusco2025}. 
At the same time, the parametrization $\epsilon_X(n_n,n_p)$ should remain sufficiently flexible to reproduce, at least approximately, existing realistic (i.e., microscopically motivated and consistent with current constraints) nucleonic EoSs, while keeping the computational cost low enough to allow large-scale sampling. 
Different implementations of $\epsilon_X(n_n,n_p)$ may emphasize these requirements to different degrees. 
The original metamodel scheme \cite{MargueronMetaI} has already proven to be appropriate for inferences of global NS observables, while also enjoying a very convenient exact mapping between some of its parameters $X$ and NMPs (discussed in Sec.~\ref{sec_map}). For these reasons, in this work we focus on this already validated framework, and improve its causal properties. We will see that this also extends its ability to reconstruct realistic EoSs over a wider range of densities and compositions.

\subsection{Mapping to the NMPs and isospin-symmetry breaking}
\label{sec_map}

Following \citet{MargueronMetaI}, it is possible to fix the behavior around saturation of $\epsilon_X$ by Taylor-expanding \eqref{eq_eX} at $n=n_0$ and $\delta = 0$ and matching it to the phenomenological expression,
\begin{equation}
\label{eq_snm}
\begin{split}
    e (x,& \delta)  =  E_0 + L_0 x + \frac{1}{2} K_0 x^2 +O(x^3)+
    \\
    +& \delta \left( E_1 + L_1 x + \frac{1}{2} K_1 x^2 +O(x^3)\right)+
    \\
    +& \delta^2 \left( E_2 + L_2 x + \frac{1}{2} K_2 x^2 +O(x^3) \right) + O(\delta^3)
\end{split}
\end{equation}
where the coefficients $\chi_1$ (with $ \chi_i=\{  E_i, L_i, K_i \}$) are typically small in absolute value compared with $\chi_0$ and $\chi_2$ \citep{Haensel_1977}. The coefficient $L_0$, which belongs to $\chi_0$, is exactly zero by construction.\footnote{The so-called saturation density $n_0$ is defined to be the density at which~$L_0=0$.} 
In the above expansion, the dimensionless variable
\begin{equation}
\label{eq_xvergognoso}
x= (n-n_0)/(3 n_0)    
\end{equation}
has been used, so that the $\chi_0$ and $\chi_2$ parameters correspond to the usual isoscalar and isovector NMPs, respectively~\citep{Chabanat1997,dutra12,rocamaza2018PrPNP,Vidana2019nqn,Drischler_2021,Grams22PRC}.

Similarly, we can fix $\delta=1$ and expand PNM around $n_0$~\citep{Somasundaram_2021},
\begin{equation}
\label{eq_pnm}
    e (x, 1)  = \tilde{E} + \tilde{L} x + \frac{1}{2} \tilde{K} x^2 + O(x^3) \, .
\end{equation}
Since no fundamental thermodynamic feature of PNM occurs specifically at $n_0$, the above expression is used only as a reference to tune $e_X$ around the benchmark density $n_0$. The coefficients $\tilde{\chi}=\{ \tilde{E},\tilde{L},\tilde{K} \}$ are not directly constrained by nuclear physics experiments but may be extracted from theoretical computations of~PNM. 

Phenomenological models typically consider only even powers of $\delta$, as required if isospin is an exact symmetry, and indicate a small but nonzero quartic contribution that is not fully exhausted by the Fermi gas term \citep{Kaiser_2015}.
However, the symmetry $\epsilon(n,\delta) = \epsilon(n,-\delta)$ is explicitly broken whenever different $m_n$ and $m_p$ are used.
Consequently, in~\eqref{eq_snm}, small nonzero coefficients $\chi_1$ are to be expected due to charge-symmetry breaking~\citep{Haensel_1977,Miller_1995,Huth2021}.

\section{Asymptotically causal metamodel}
\label{sec_eX024}

We implement a minimal scenario in which isospin symmetry, $\epsilon(n,\delta) = \epsilon(n,-\delta)$, is broken only by the physical mass imbalance, and therefore retain only the terms corresponding to the lowest even powers of $\delta$ in the metamodel ansatz~\eqref{eq_meta_ansatz}. Accordingly, we assume an energy per baryon of the form
\begin{equation}
\label{eq_eX024}
\begin{split}
e_X(n_n,n_p) = & e_F(n,\delta) +
    \\ &+u_0(n) + u_2(n)\delta^2 + u_4(n)\delta^4 \, ,        
\end{split}
\end{equation}
where 
\begin{equation}
\label{eq_eF024}
 e_F(n,\delta)= \frac{1}{n}\sum_q\left( \epsilon_F^q(n_q)-m_q n_q \right) .
\end{equation}
In the following, we motivate and specify the explicit parametrization adopted for~$u_i(n)$.

\subsection{Asymptotic causality}

The original formulation in \citep{MargueronMetaI} adopts a specific parametrization of the functions~$u_i(n)$:
\begin{equation}
    u_0(n)=P_0(x),\quad u_2(n)=P_2(x), \quad u_4(n)=0 \, ,
\end{equation}
where $P_0(x)$ and $P_2(x)$ are 4th-order polynomials in $x$, multiplied by a low-density correction that enforces $u_i(n=0)= 0$; see equation (37) therein. 
This low-density correction is negligible around and above saturation, so we will consider just the polynomial nature of $P_i(x)$ in the following. 
The polynomial form of $P_i(x)$ is motivated by analytical convenience, in particular by the possibility of obtaining a simple mapping between some of the original metamodel parameters $X$ and the NMPs. 
The choice of 4th-order $P_i(x)$ is further required to provide sufficient flexibility to reproduce the energy per baryon of SNM and PNM of popular nuclear models~\citep{MargueronMetaI,MargueronMetaII}.

However, this choice also entails an intrinsic drawback: since $P_0(x)\sim P_2(x)\sim (n/n_0)^4$ for $n\gg n_0$, the resulting sound speed $v$ must grow toward the asymptotic value $v^2=4$, and is therefore doomed to become superluminal at sufficiently high density.\footnote{
    Assume $\epsilon = a n^\gamma + m n $, for $\gamma>1$. Then, $P = (\gamma-1) a n^\gamma$ and $v^2= d P / d \epsilon$ approaches the asymptotic value $\gamma-1$. This asymptotic value $v^2 \approx \gamma-1$ is attained at high densities ($ \gamma a n^{\gamma-1} \gg m $) and overshoots unity if~$\gamma >2$.
    } 
In practice, this problem is addressed by extensively exploring the parameter space associated with $P_0$ and $P_2$ in order to identify metamodel instances that remain causal and stable in the $(n,\delta)$ regime relevant to NS cores. 
Such a procedure, however, entails a significant computational overhead and represents a limitation of the original metamodel, typically leading to a high rejection rate in Bayesian studies. 
This shortcoming can be mitigated by adopting a different form for the $u_i$ that satisfies asymptotic causality: the $u_i$ have to grow no faster than $u_i\sim n$, guaranteeing that $\epsilon_X \sim n^2$ at most (i.e., $v^2\leq 1$ asymptotically)~\citep{Zeldovich1961}.

A related strategy was introduced in~\citep{Huth2021}, where the asymptotic scaling $\epsilon_X \sim n^{5/3}$ was chosen so as to improve the high-density stability properties of the model by recovering the non-relativistic Fermi-gas scaling.
This alternative metamodel construction was later employed~\citep{Lim2024}, and is well suited to enforcing asymptotic causality~($v^2 = 2/3$ asymptotically).

Clearly, the asymptotic requirement alone does not guarantee the causality and stability of the resulting $e_X$ over the whole $(n,\delta)$ domain above saturation. However, we find that this helps achieve substantially lower rejection rates even when the model is considered up to realistic central densities of NSs; i.e., up to~$\sim 1.2\,$fm$^{-3}$.

\subsection{An asymptotically causal parametrization}

The form we assume for $u_i$ ($i=0,2$) is
\begin{equation}
\label{eq_u02}
    u_i(n) = V_i(x) + 
    \frac{h^{(0)}_i + h^{(1)}_i x + h^{(2)}_i x^2 + h^{(3)}_i x^{3}}
    {(1 + a_i x)(1 + b_i x)(1 + c_i x)} \, ,
\end{equation}
where the low-density correction $V_i$ ensures that $u_i(n)\propto n$ for $n\ll n_0$ without affecting the NMP mapping at saturation or the asymptotic sound speed at high density. 
In the above expression, the parameters $h_i^{(j)}$ for $j=0,\ldots,3$ have the physical dimensions of an energy per baryon, while 
\begin{equation}
    0\leq a_i < b_i \leq c_i<3
\end{equation}
are dimensionless (values very close to 3 are excluded to avoid a close-to-zero denominator when $n\rightarrow 0$).
The specific choice in \eqref{eq_u02} can be motivated as follows:
\begin{enumerate}
\item For $x\gg1$, the numerator grows as $n^3$. The denominator ensures asymptotic causality by lowering the growth to at most $u_i\sim n$ in the limiting case $a_i=0$ ($u_i\sim \mathrm{const}$ for $a_i>0$). 
\item For suitable choices of $V_i$, it allows an exact mapping between the NMPs and the unknown $h^{(j)}_i$, as shown in App.~\ref{app_mapping}.
\item The $u_i(n)$, if interpreted as the total energy per baryon, give an asymptotic (frozen) sound speed of $v=0$ in the typical case $a_i>0$. This means that, with the inclusion of the Fermi gas term $e_F$ in \eqref{eq_eX024}, the asymptotic sound speed calculated from $e_X$ is $v=1/\sqrt{3}$. This is different from some relativistic matter models, where the asymptotic sound speed is expected to saturate to 1 \citep{Zeldovich1961,alford2022PhRvC}. Since it is possible that the sound speed overshoots its asymptotic value at intermediate densities of a few times $n_0$, this lower choice for the asymptotic sound speed (i.e., $1/\sqrt{3}$ rather than $1$) helps us explore a wider range of sound-speed behaviors while remaining within the causality and stability bounds over the density range relevant to NSs. 
\end{enumerate}
The remaining term $V_i(x)$ must guarantee reasonable subsaturation behavior without spoiling the aforementioned mapping to the NMPs. Several choices are possible, but we find it convenient to use
\begin{equation}
\label{eq_V02}
    V_i(x) = 
    \frac{s_i\, |3x|^{3+g_i}}{1 + w_i(3x+1)^{3+g_i}},  
\end{equation}
where   
$w_i>0$, $g_i \geq 0$, and
\begin{equation}
\label{eq_s02}
    s_i =
    -\frac{27 h_i^{(0)} - 9  h_i^{(1)} + 3  h_i^{(2)} -  h_i^{(3)}}
    {(3 - a_i)(3 - b_i)(3 - c_i)} \, .
\end{equation}
This guarantees that $u_i(0)=0$.

The expression in \eqref{eq_V02} allows us to explore different low-density behaviors through the parameters $w_i$ and~$g_i$. While the condition $w_i>0$ keeps the denominator positive for all $n\geq0$ and makes $V_i$ asymptotically constant, the requirement $g_i \geq 0$ ensures continuity of the sound speed at saturation; the stricter condition $g_i>0$ also ensures continuity of its first derivative, as discussed in App.~\ref{app_g}.

{
It is important to stress that our form \eqref{eq_u02} is by no means unique. Similar to other agnostic parametrization schemes (e.g., spectral representations, piecewise polytropes, sound-speed parametrizations), it should not be seen as a microscopic ansatz but only as a controlled composition-aware functional prior. Alternative composition-aware schemes (e.g., \citep{MargueronMetaI,Lim2024}) remain necessary cross-checks, exactly as alternative barotropic schemes are currently used to assess systematic uncertainty in agnostic barotropic EoS inference~\citep{Rutherford_2024,biswasPRD2026}.
}

\subsection{Quartic Correction in $\delta$}
\label{sec_quartcorr}

The Taylor expansion of $e_X$ in \eqref{eq_eX024} around $\delta=0$ contains an infinite number of both odd and even powers of $\delta$ at any density, including saturation. This occurs simply because $e_F(n,\delta)$ is not a polynomial in $\delta$, and we use different physical masses $m_n$ and~$m_p$.

Therefore, the general scheme in \eqref{eq_eX024}, which also includes the original metamodel \citep{MargueronMetaI}, goes beyond the parabolic approximation (i.e., $e(n,\delta) = e_0(n) + \delta^2 e_2(n)$) by construction.
However, unless an effective phenomenological mass is introduced in $e_F(n,\delta)$, the kinetic term $e_F$ is identical across all nuclear models. As a result, limiting the expansion to $u_0$ and $u_2$ forces all higher-order terms in $\delta$ to be identical for every model. 
This may prevent the reproduction of different realistic nuclear models, especially in their PNM sector ($\delta=1$). To retain minimal complexity while increasing flexibility, we therefore include the next relevant even term \cite{Seif_2014,Cai_2015,Kaiser_2015},
\begin{equation}
\label{eq_u4}
    u_4(n) = A\,\frac{n/n_0}{1 + (n/n_0)^B} \, ,
\end{equation}
where $A$ and $B>0$ raise the total number of parameters in $X$ to~21.

The correction \eqref{eq_u4} introduces additional flexibility beyond the quadratic expansion without violating isospin symmetry or spoiling the mapping to nuclear matter parameters. It is designed to preserve the separation between the symmetry energy, $e_{\text{sym}}$, and the pure neutron matter (PNM) energy. Specifically, since the symmetry energy is usually defined as
\begin{equation}
    e_{\text{sym}} = \left.\frac{1}{2}\frac{\partial^2 e}{\partial \delta^2}\right|_{x=0,\, \delta=0},
\end{equation}
a purely quadratic expansion, i.e., with only $u_0$ and $u_2$, can correctly reproduce the symmetry energy near saturation density and at small asymmetry ($\delta \approx 0$) for a given EoS, but could fail to accurately describe the pure neutron matter energy of the same model, which corresponds to~$\delta = 1$.

The inclusion of $u_4(n)$ allows us to better match the PNM energy without altering the behavior around symmetric matter. The parameters $A$ and $B$ can either be determined from \eqref{eq_pnm}, if known, or by fitting the PNM energy around saturation density $n_0$, as done in Sec.~\ref{sec_fits}.

\subsection{Parameters}

The two functions $u_i$ in \eqref{eq_u02} contain 19 free parameters: $h^{(0)}_i$, $h^{(1)}_i$, $h^{(2)}_i$, $h^{(3)}_i$, $a_i$, $b_i$, $c_i$, $g_i$, $w_i$, for $i=0,2$, plus $n_0$, which is hidden in $x$ through the definition \eqref{eq_xvergognoso}. The term $u_4$ contains two additional parameters, bringing the total number of parameters in $e_X(n,\delta)$ to 21.
However, the expansion of $e_X(n,\delta)$ around saturation, see \eqref{eq_snm}, does not guarantee that $L_0=0$. Imposing this constraint reduces the total number of free parameters by one. To this end, we implement an exact mapping between the parameters $h^{(0)}_i$, $h^{(1)}_i$, $h^{(2)}_i$ and the low-order NMPs, as detailed in App.~\ref{app_mapping}. 

Thanks to this mapping, the complete set of parameters in $e_X(n,\delta)$ is:
\begin{equation}
\label{eq_califfo_bidone}
    \{ E_i, K_i, n_0, L_2 , w_i, g_i, h_i^{(3)}, a_i, b_i, c_i , A, B  \} 
\end{equation}
for $i=0,2$, namely 20 parameters (18 if $u_4$ is suppressed).
The first six parameters $\{ E_i, K_i, n_0, L_2 \}$ in \eqref{eq_califfo_bidone} are considered physical and are therefore directly sampled using the available nuclear-physics information on the low-order NMPs.
The remaining 14 parameters are not directly constrained by empirical input and have no known connection to nuclear observables. They are therefore only by-products of our chosen parametrization.

It is important to note that, with the exception of the six NMPs $ \{ E_i, K_i, n_0, L_2\}$, the parameters in \eqref{eq_califfo_bidone} should be interpreted as parametrization coordinates rather than physical quantities, similarly to polytropic indices, spectral coefficients, Gaussian-process latent variables, or sound-speed nodes in agnostic models of the barotropic EoS. 

Even though $\{  w_i, g_i, h_i^{(3)}, a_i, b_i, c_i , A, B  \}$ are subject to physical bounds imposed by stability and causality, as well as analytical constraints arising from the specific parametrization, we expect \emph{a priori} that their posterior distributions will remain largely prior dominated and exhibit a certain degree of practical non-identifiability and approximate redundancy. The full parametrization  $e_X$  nevertheless provides the flexibility required for an accurate reconstruction of specific microscopic EoSs, as will be shown in Sec.~\ref{sec_fits}. For the Bayesian exploration, however, some parameter directions may be predictively redundant: they can be fixed without loss of effective model coverage, provided that this does not appreciably modify the induced prior and posterior predictive distributions of the physical observables or the posteriors of the retained parameters~(see Sec.~\ref{sec_priora}).

\subsection{Neutron star matter in the nucleonic hypothesis}
\label{sec_NSmatt}

For NS applications, the cold nuclear matter model $\epsilon_X(n_n,n_p)$ can be minimally complemented to include electrons ($e$) and muons ($\mu$)~\citep{ShapiroBook,HaenselBOOK2007}:
\begin{equation}
\label{eq_eps_tot}
\begin{split}
\epsilon^{tot}_X(n_i) = \epsilon_X(n_n,n_p)+\sum_{l=e,\mu} \epsilon_F^l(n_l) \, ,
\\
P^{tot}_X(n_i) = P_X(n_n,n_p)+\sum_{l=e,\mu} P_F^l(n_l) \, ,
\end{split}
\end{equation}
where $i=n,p,e,\mu$, while $\epsilon_F^l$ is given in \eqref{eq_eps_F} and $P_F^l$ in~\eqref{eq_PF}. 
While \eqref{eq_eps_tot} in principle also describes configurations outside $\beta$-equilibrium and charge neutrality, for most NS applications we will use its electrically neutral, chemically equilibrated section,
\begin{equation}
\label{eq_eps_totbeta}
\epsilon^{\beta}_X(n) = \epsilon^{tot}_X\left(n x^\beta_i(n) \right)  \, ,
\end{equation}
where $n$ is the baryon number density and $x_i^\beta(n)$ ($i=n,p,e,\mu$) are the species fractions at $\beta$-equilibrium.\footnote{
    They satisfy the usual constraints $0\leq x_i^\beta(n)\leq1$, $x_n^\beta(n)+ x_p^\beta(n)=1$, and $x_p^\beta(n) = x_e^\beta(n) + x_\mu^\beta(n)$.
    }
It follows that the pressure along the $\beta$-equilibrated section is
\begin{multline}
\label{eq_Pbeta}
P^\beta_X(n) = n^2 \dfrac{d}{dn} \dfrac{\epsilon^{\beta}_X(n)}{n} = 
P^{tot}_X\left(n \, x^\beta_i(n) \right) \, ,
\end{multline}
where $P_X$ is given in \eqref{eq_PXeuler} and $P_F^l$ in~\eqref{eq_PF}. 

A measure of the stiffness of the EoS $\epsilon^{\beta}_X(n)$ is its equilibrated squared sound speed,
\begin{equation}
\label{eq_vs_beta}
    v_{\beta X}^2(n) = \frac{d P^\beta_X(n)/dn}{d\epsilon^\beta_X(n)/dn} \, .
\end{equation}
However, physical sound signals propagate at the speed
\begin{equation}
\label{eq_vs_FR}
    v_{fX}^2(n) = \left.
    \frac{d P^{tot}_X(n x_i)/dn}{d\epsilon^{tot}_X(n x_i)/dn} 
    \right|_{x_i= x^\beta_i(n)} \, ,
\end{equation}
which is therefore the quantity to be checked when demanding special-relativistic causality~\citep{camelio_I}. We refer to this quantity as the frozen (i.e., fixed-composition) sound speed, the chemical analog of the more common adiabatic sound speed; see, e.g.,~\citep{HaenselAdiab,Friedman_2017}. 
In particular, if cold-catalyzed matter is both stable and causal close to equilibrium, it follows that $0<v_{\beta X}^2(n)<v_{fX}^2(n)<1$~\citep{camelio_I,Montefusco2025}.

This treatment is valid in the core of an NS, where one expects cold-catalyzed homogeneous nuclear matter. For the outer layers, namely the inner and outer crust \citep{Chamel2008LRR}, our starting point remains the minimal extension in \eqref{eq_eps_tot}, but instead of solving for the $\beta$-equilibrium of homogeneous matter to obtain the $x_i^\beta(n)$, we adopt the compressible liquid-drop model approach presented in~\citep{Carreau_2019,DinhThi_2021a,davis2024}. 
This approach remains consistent with more advanced extended Thomas-Fermi calculations \citep{Grams_2022,klausner2025prc,grams_diverres_2025}, and provides a convenient procedure to compute the thermodynamic and composition properties of the solid crust, as well as the crust-core transition, within a unified model based on the same $e_X$ in~\eqref{eq_eX024}.

\section{Fits to realistic equations of state}
\label{sec_fits}

Ideally, the metamodel in \eqref{eq_eX024} should be flexible enough to reproduce realistic nuclear EoSs, namely EoSs that are both compatible with current astrophysical constraints and microscopically grounded~\citep{fortin_and_everyone_2016,oertel_RMP_17,burgiofantina2018}.
Clearly, a metamodel fit to any realistic EoS cannot be exact, but aiming for an extremely precise reconstruction is unnecessary given the high level of uncertainty associated with current EoSs. However, the accuracy with which the target microscopic EoS is reproduced provides a measure of the flexibility of the metamodel scheme. On the other hand, regions where the fit is poor can be used to highlight a density or isospin regime in which the metamodel should be improved.

In view of this, we now attempt to reproduce some microscopically motivated EoSs. Since no fit is exact, even small differences between $\epsilon_X(n,\delta)$ and the target $\epsilon(n,\delta)$ can lead to larger differences in derived thermodynamic quantities, particularly the chemically equilibrated composition and the sound speed, because these depend on first and second derivatives of $\epsilon_X$. Formally, this is due to error amplification under differentiation.

To test these points, we selected five widely used models that differ significantly in their properties: SLy4 \citep{Chabanat1997,DouchinAA2001}, BSk24 \citep{goriely_prc_2013,BSk24,Pearson_2018}, DD2 \citep{typelPRC2010,hempelNPA2010}, FSU2 \citep{Chen_2014}, and TM1e \citep{ShenEoS,Shen2020ApJ}. The first two are non-relativistic Skyrme-type functionals, while the latter three are based on RMF models.

\subsection{Fit procedure for the EoS reconstruction}

{ We use the full 20-parameter representation in \eqref{eq_califfo_bidone} to reproduce the selected microscopic models. Given the large number of parameters, we develop a fitting strategy dictated by the functional form~\eqref{eq_eX024}. }

For $\delta=0$, only the parameters entering $u_0(n)$ contribute. We therefore determine $u_0(n)$ from the symmetric nuclear matter (SNM) energy per baryon of a given target EoS, and subsequently fix the isovector sector, i.e., $u_2(n)$ and $u_4(n)$, from the pure neutron matter (PNM) slice. Although this procedure uses only the two slices $(n,\delta=0)$ and $(n,\delta=1)$ as input, the ansatz in \eqref{eq_eX024} provides an explicit extension to arbitrary $(n,\delta)$: once the parameters are fixed, $\epsilon_X(n,\delta)$ is defined for all asymmetries. Hence, the fit defines an EoS reconstruction procedure, because it determines a full two-dimensional energy landscape from the two fitted slices, which can then be tested a posteriori at intermediate compositions.

We validate the reconstruction on the cold $\beta$-equilibrated EoS obtained by solving the chemical-equilibrium conditions after augmenting the metamodel with leptons (Sec.~\ref{sec_NSmatt}). Agreement in $\beta$-equilibrium is not guaranteed a priori, because it probes the interpolation of \eqref{eq_eX024} at intermediate asymmetry and involves first and second derivatives of $\epsilon_X(n,\delta)$, which amplify small residuals in the fitted energy. We quantify the reconstruction accuracy by comparing, in cold $\beta$-equilibrium, the energy density, pressure, sound speed, and particle fractions for the metamodel and the target EoS.

For each model, the saturation density and the nuclear matter parameters (NMPs) up to second order are taken from the CompOSE~\citep{compOSE} tables and kept fixed in the fit. The quartic-correction coefficients $A$ and $B$ are mapped to $\tilde{E}$ and $\tilde{L}$ defined in \eqref{eq_pnm}, thereby enforcing the PNM expansion around saturation up to first order (see App.~\ref{app_mapping}). 
{ The remaining 12 parameters are treated as free and optimized. }

\subsection{EoS reconstruction: results}

\begin{figure}
\centering
\includegraphics[width=\columnwidth]{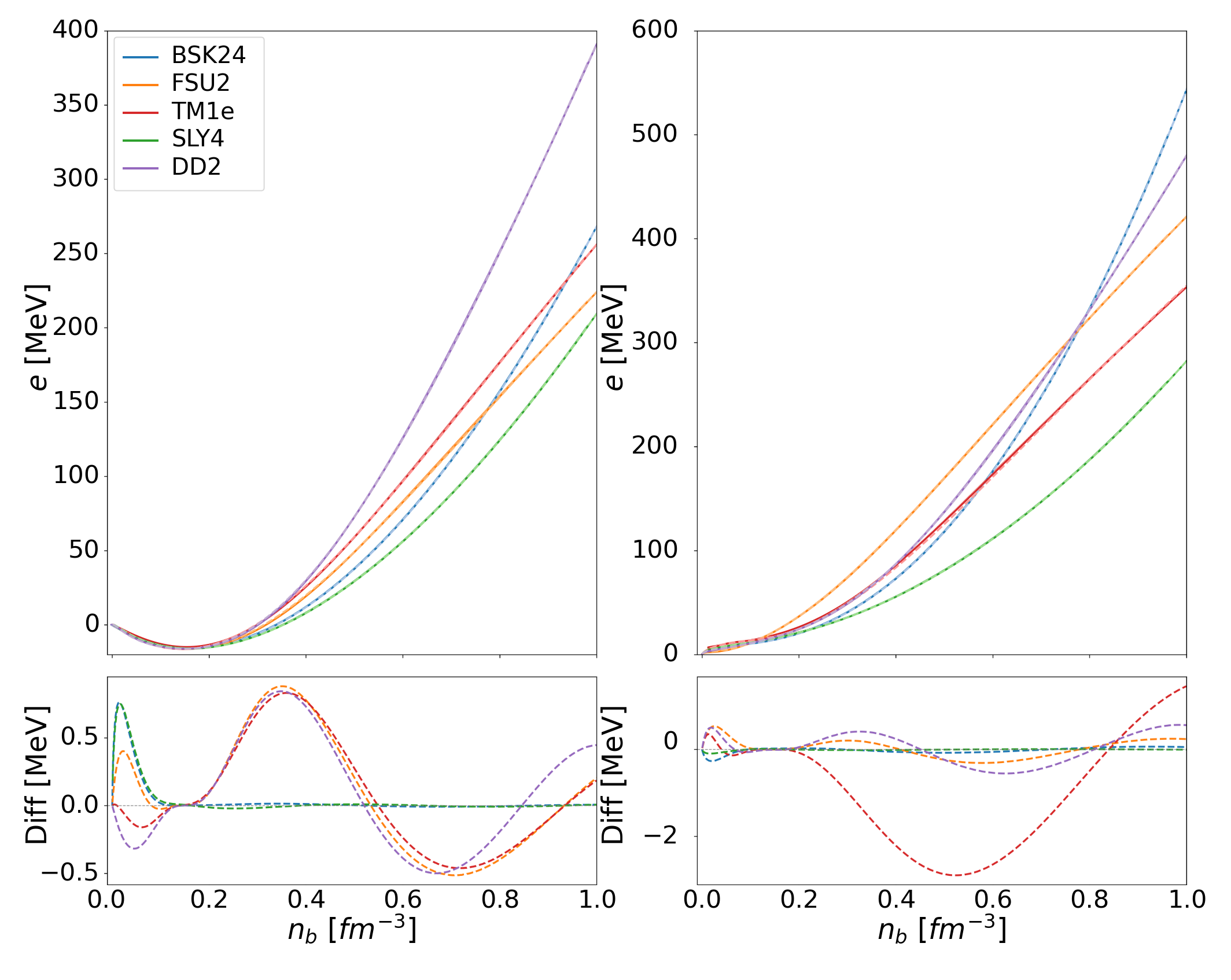}
\caption{
    Comparison of fitted and original energy per baryon.
    Top panels show the original and fitted $e_X(n,\delta)$ for SNM ($\delta=0$, left) and PNM ($\delta=1$, right); bottom panels show the differences between the original and fitted models in~MeV.
}\label{fig_eos_comparison}
\end{figure}

The accuracy of the reconstruction for the SNM and PNM energy per baryon is shown in Fig.~\ref{fig_eos_comparison}. In both panels, the residual vanishes at saturation by construction: on the SNM side this follows from the mapping to the empirical NMPs at $n_0$, while for PNM it is enforced by the quartic correction through $u_4$ around saturation (see App.~\ref{app_mapping}). 

Above saturation, the reconstruction remains very accurate for all tested EoSs, with relative deviations that stay small and typically decrease with increasing density. 

At subsaturation densities, the reconstruction remains reasonable, but the fit quality degrades, suggesting that the low-density corrections $V_i$ are not flexible enough to maintain the same level of accuracy. This is not surprising, as it is known that the low-density regime requires well-tuned corrections~\citep{vidana_low_21,Grams:2024hdl,burrello2025crust}. However, this regime mainly affects the crust \citep{Grams2021,burrello2025crust,klausner2025prc}, where PNM is the relevant component at very low densities, while the ionic contribution is controlled by the energy behavior close to saturation. Moreover, this part of the EoS is more relevant for crust-sensitive dynamical phenomena (e.g., cooling and pulsar glitches \citep{Chamel2008LRR,AMP_arxiv_2023}), which are beyond the scope of the present work.

\begin{figure*}[t]
    \centering
    \begin{minipage}[t]{0.45\textwidth}
        \centering
        \includegraphics[width=\linewidth]{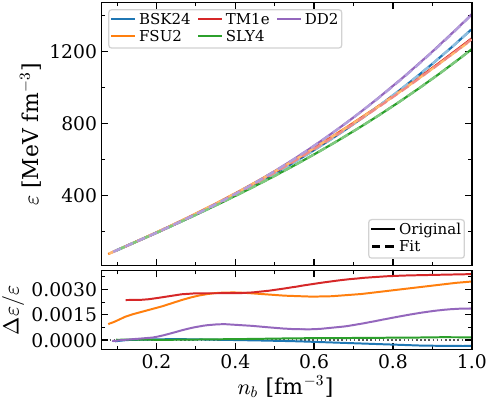}
        
        \small (a) Energy Density
    \end{minipage}
    \hfill
    \begin{minipage}[t]{0.45\textwidth}
        \centering
        \includegraphics[width=\linewidth]{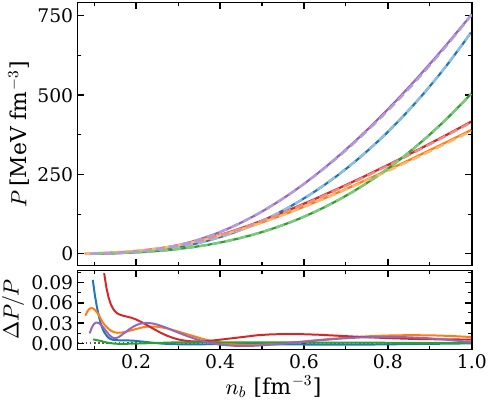}
        
        \small (b) Pressure
    \end{minipage}

    \vspace{0.5cm}
    \begin{minipage}[t]{0.45\textwidth}
        \centering
        \includegraphics[width=\linewidth]{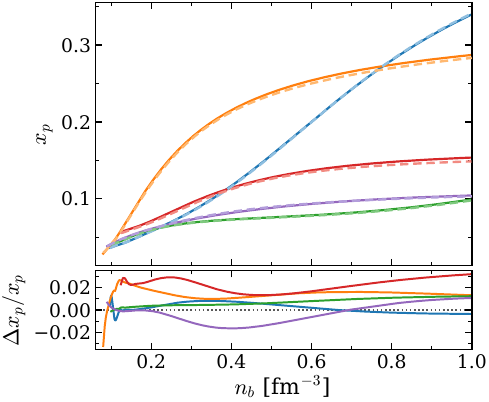}
        
        \small (c) Composition
    \end{minipage}
    \hfill
    \begin{minipage}[t]{0.45\textwidth}
        \centering
        \includegraphics[width=\linewidth]{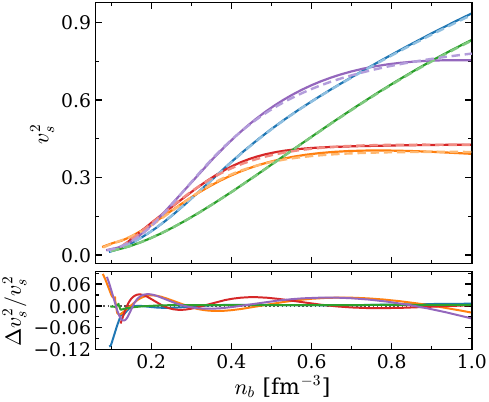}
        
        \small (d) Speed of Sound
    \end{minipage}

    \caption{Comparison of physical quantities across models: (a) energy density, (b) pressure, (c) proton fraction, and (d) equilibrated speed of sound. All quantities are calculated over the barotropic slice defined by $\beta$-equilibrium. }
    \label{fig_4panel_comparison}
\end{figure*}

As we have already stressed, reproducing the SNM and PNM slices does not, in principle, guarantee an accurate reconstruction of the cold $\beta$-equilibrated EoS. Nonetheless, for the EoSs considered here the reconstructed energy landscape is sufficiently accurate that the resulting barotropic EoS in $\beta$-equilibrium agrees with the target within a few percent, as shown in Fig.~\ref{fig_4panel_comparison}. In that figure, solid lines denote the metamodel reconstruction, while dashed lines denote the original CompOSE EoS. Since our focus is on homogeneous matter, the comparison starts at the crust-core transition predicted by our CLDM for each EoS. Over the core-density range, not only the pressure and energy density but also the sound speed and composition are typically reproduced within $5\%$, with the largest deviations confined to a narrow region close to the crust-core transition.

\section{Probing the metamodel's parameter space}
\label{sec_par_space}

We now turn to the flexibility of $e_X$ in exploring the EoS space. To do this, we perform a Bayesian analysis that follows the methodology described in \citet{Montefusco2025}, while incorporating the latest NICER observations of PSR J0614$-$3329 \citep{Nicer2025} and the Shapiro delay measurement of the massive pulsar PSR J0740+6620~\citep{J0740Mass}.

{ Since the presence of a phase transition cannot be excluded on the basis of recent agnostic Bayesian inferences \citep{Brandes_2023,Brandes_2025}, we adopt the nucleonic hypothesis here as a controlled baseline, while phase transitions and exotic degrees of freedom remain viable scenarios outside the scope of the present work. In future studies, the nucleonic metamodel could be augmented with transition parameters, and the resulting model class could be compared with the nucleonic one, as done in~\cite{mondal2023MNRAS}. 
}

Unlike previous studies based on the original metamodel (e.g., \citep{MargueronMetaII,CarreauMeta,guven20,HoaUniverse,davis2024,ZhangMeta,klausner2025prc,burrello2025crust,Koehn_2025,Klausner_prc_2026,diverres_shear_2026}), here we implement the more restrictive stability-causality criterion discussed in~\citep{camelio_I,Montefusco2025}, 
\begin{equation}
\label{eq_caus_stab}
0<v_{\beta X}^2(n)<v_{fX}^2(n)<1    
\end{equation}
for the $\beta$-equilibrated and adiabatic (frozen) sound speeds, see \eqref{eq_vs_beta} and \eqref{eq_vs_FR}. 
Note that, in inferences based on agnostic sound speed models that only describe the barotropic sector of the EoS (e.g., \citep{altiparmak2022ApJ,Brandes_sound_inference_23,Brandes_2023,Brandes_2025}), the causality condition is necessarily the slightly weaker one $0<v_{\beta X}^2<1$; the possibility of including the stronger physical check in \eqref{eq_caus_stab} is an extra advantage of using a composition-aware formalism~\citep{Montefusco2025}.

\subsection{Prior}
\label{sec_priora}

{
The prior $P(X)$ for the parameters $X$ of $e_X$ listed in \eqref{eq_califfo_bidone} is constructed as follows. While the full 20-parameter representation is retained for the accurate reconstruction of specific microscopic EoSs in Sec.~\ref{sec_fits}, for the Bayesian exploration we set $g_i=0$ and $u_4=0$ (implemented by setting $A=0$; $B$ is then irrelevant and is set to zero by convention). This reduction is justified empirically: we ran an exploratory analysis with these parameters allowed to vary over broad domains and found that they do not appreciably enlarge the induced posterior predictive distributions of the EoS curves and physical observables, nor modify the marginalized posteriors of the retained parameters. For the present likelihood, they therefore correspond to practically non-identifiable directions. Fixing them reduces the dimensionality of the sampling problem and improves its efficiency without appreciably reducing the effective predictive coverage of the model. The remaining 16 parameters
\begin{equation}
      \{ E_i, K_i, n_0, L_2 , w_i, h_i^{(3)}, a_i, b_i, c_i \}
\end{equation}
are sampled as follows.
}

{
For $\{ w_i, h_i^{(3)} \}$, we begin with intervals encompassing the minimum and maximum values obtained from the fits discussed in Sec.~\ref{sec_fits}. These bounds are then broadened until both the posteriors of the retained parameters and the posterior predictive distributions become insensitive to further changes, indicating that the flat prior is sufficiently wide. 
The prior on $\{a_i,b_i,c_i \}$ is discussed in App.~\ref{app_abc}.
Finally, the NMPs $\{ E_i, K_i, n_0, L_2\}$ are sampled from the prior distributions defined in \cite{Carreau_2019}, which have been used in essentially all previous metamodel studies. 
}

The details of the resulting full prior are reported in Tab.~\ref{tab_prior-bounds}.

\subsection{Bayesian procedure}

The prior $P(X)$ is updated using Bayes' theorem with a likelihood that assigns a weight $\mathcal{L}_D(X)$ to each $e_X$ according to the following data~$D$:
\begin{enumerate}
\item The PNM slice $e_X(n,\delta=1)$ must be consistent with the energy per nucleon of PNM, as determined from chiral effective field theory ($\chi_{EFT}$) calculations~\citep{Machleidt2016PhyS,Huth2021}. The combination of various $\chi_{EFT}$ results yields a reliable energy band that is used to build an informed prior; see Sec.~3.1 and App.~A of \citep{Montefusco2025}.
\\
\item The energy landscape $e_X(n,\delta)$ should be consistent with the nuclear mass measurements reported in the AME2020 mass evaluation~\citep{AME2020}, see Sec.~3 of~\citep{Montefusco2025}. 
\\
\item The resulting EoS $\epsilon^\beta_X(n)$ must support a maximum NS mass $M_{\mathrm{TOV}}(X)$ greater than that of PSR J0348+0432 \citep{antoniadis2013} and PSR J0740+6620~\citep{J0740Mass}. 
\\
\item We demand positive pressure, the dominant-energy condition $0<P^\beta_X(n)<\epsilon^\beta_X(n)$, and the stability-causality condition for reacting mixtures $0<v_{\beta X}^2(n)<v_{f X}^2(n)<1$ for all baryon densities in the range $0<n<n_{X}$, where $n_{X}$ is the central density corresponding to the NS configuration of mass~$M_{\mathrm{TOV}}(X)$.
\\
\item Each sampled instance $\epsilon^\beta_X(n)$ must be compatible with constraints on tidal deformability inferred from the binary NS merger event GW170817 \citep{GW170817_1}; see App.~B of \citet{Montefusco2025} for details. 
\\
\item The mass-radius estimates obtained from X-ray pulse-profile modeling by NICER for the pulsars PSR J0030+0451 \citep{NicerJ0030}, PSR J0437-4715 \citep{NicerJ0437}, PSR J0740+6620 \citep{NicerJ0740}, and PSR J0614$-$3329 \citep{Nicer2025} must be reproduced.
\end{enumerate}
The first point, namely consistency with $\chi_{EFT}$ calculations of nuclear matter, is implemented using Metropolis--Hastings sampling. We then randomly extract $10^6$ models from this informed prior and pass them through the sequence of Bayesian filters 2--6, each of which assigns a partial likelihood.

Each of the above points contributes to the total likelihood $\mathcal{L}_D(X)$ for $e_X$ in a multiplicative fashion. We refer to \citep{Montefusco2025} for a detailed description of our Bayesian procedure and the explicit implementation of each likelihood factor.

\begin{table}[ht]
\centering
\caption{
    {
    Prior bounds for the 16 parameters used in the Bayesian analysis. The prior on $\{a_i,b_i,c_i\}$ is not flat, and is discussed in App.~\ref{app_abc}. As discussed in the text, $A$, $B$, and $g_i$ are set to zero.
    }
} \label{tab_prior-bounds}
\begin{tabular}{lcc}
\hline
$\qquad$ $\qquad$ & $\qquad$min$\qquad$ & $\qquad$max$\qquad$ \\
\hline
$n_0$ & 0.15 & 0.17 \\
$E_0$ & -17 & -15 \\
$K_0$ & 190 & 270 \\
$w_0$ & $0^+$ & 10 \\
$h_0^{(3)}$ & 0 & 300 \\
$E_2$ & 22 & 38 \\
$L_2$ & -20 & 125 \\
$K_2$ & -1000 & 1000 \\
$w_2$ & $0^+$ & 10 \\
$h_2^{(3)}$ & 0 & 400 \\
$a_i$ & 0 & $b_i$ \\
$b_i$ & $a_i $ & $c_i$ \\
$c_i$ & $b_i$ & $3^-$ \\
\hline
\end{tabular}
\end{table}

\section{Results}
\label{sec_result}

We sample the prior $P(X)$, calculate the total likelihood $\mathcal{L}_D(X)$ for each sampled instance $X$, and obtain the posterior $P(X|D) \propto \mathcal{L}_D(X) P(X)$. For each sampled instance $X$, we use the calculated $\mathcal{L}_D(X)$ to assign a weight to each quantity pertaining to the model~$e_X$ in~\eqref{eq_eX024}, where the quartic term $u_4$ is set to zero.\footnote{
    $P(X|D)$ is obtained by setting $A=B=0$ in \eqref{eq_u4} at the level of the prior $P(X)$. The term $u_4$ decouples the PNM and SNM energies around saturation, but has no appreciable impact on the posterior distributions considered here.
}

The posterior results can be categorized into two main groups: global and microscopic properties. 
Global NS properties are more directly constrained by astrophysical observations. In contrast, microscopic properties, such as the proton fraction and the speed of sound, remain largely unconstrained at the densities found in NS interiors, because $\mathcal{L}_D(X)$ contains no direct experimental information in this density range. Indeed, the laboratory and $\chi_{EFT}$ data we considered in $D$ provide constraints only up to, or slightly above, nuclear saturation density.

\begin{table}[ht]
\centering
\begin{tabular*}{\columnwidth}{@{\extracolsep{\fill}}lccccccc@{}}
\toprule
& \textbf{Units} & \textbf{Median} & \multicolumn{2}{c}{\textbf{68\% CI}} & \multicolumn{2}{c}{\textbf{95\% CI}} \\
 &  &  & \textbf{Min} & \textbf{Max} & \textbf{Min} & \textbf{Max} \\
\midrule
$M_{\mathrm{TOV}}$     & $M_\odot$   & 2.25    & 2.15    & 2.38    & 2.07    & 2.56    \\
$M_{\mathrm{dU}}$      & $M_\odot$   & 1.71    & 1.15    & 2.18    & 0.89     & 2.50    \\
$R_{1.0}$              & km          & 12.05   & 11.50   & 12.54   & 10.99   & 12.93   \\
$R_{1.4}$              & km          & 12.21   & 11.68   & 12.61   & 11.20   & 12.99   \\
$R_{2.0}$              & km          & 12.09   & 11.48   & 12.57   & 10.57   & 13.02   \\
$\Lambda_{1.0}$        & ---         & 2743    & 2052    & 3384    & 1563    & 4108    \\
$\Lambda_{1.4}$        & ---         & 436     & 321     & 533     & 245     & 664     \\
$\Lambda_{2.0}$        & ---         & 37      & 24      & 52      & 10      & 70      \\
$f_{1.0}$              & kHz         & 1.57    & 1.48    & 1.69    & 1.40    & 1.81    \\
$f_{1.4}$              & kHz         & 1.79    & 1.72    & 1.91    & 1.65    & 2.01    \\
$f_{2.0}$              & kHz         & 2.02    & 1.93    & 2.15    & 1.84    & 2.38    \\
\bottomrule
\end{tabular*}
\caption{
Posterior medians and 68\% and 95\% credible intervals for selected global stellar quantities; see text for details.
}
\label{tab:stellar_quantities}
\end{table}

\subsection{Composition-independent global NS properties}

\begin{figure}[htbp]
    \centering
    \includegraphics[width=\columnwidth]{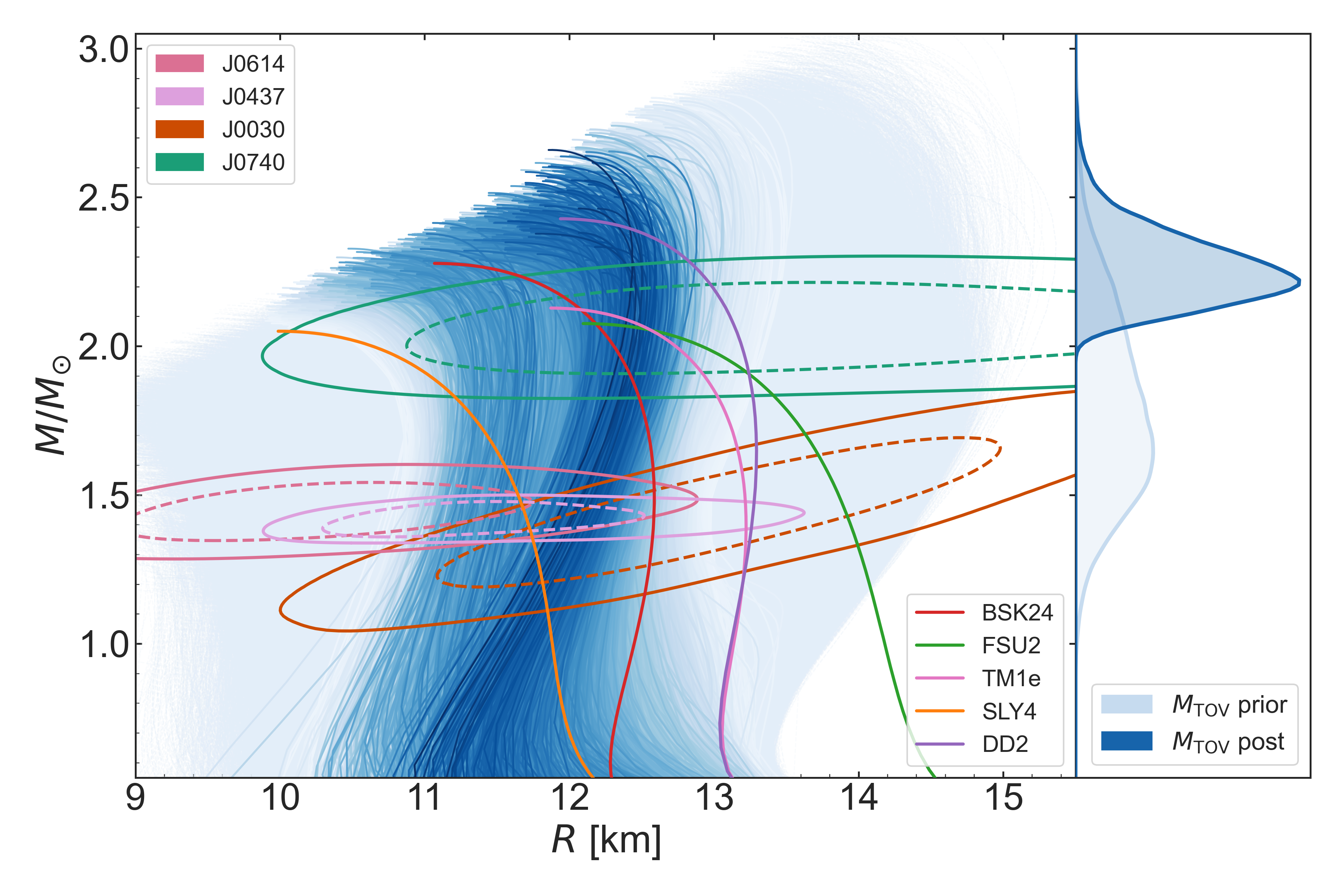}
    \caption{
    Mass-radius relations $R_X(M)$ for cold non-rotating NSs obtained from the sampled models $e_X$. Models that fail the stability-causality requirement up to $M_{\mathrm{TOV}}$ are excluded. The light gray background shows the informed prior, corresponding to displayed models with $\mathcal{L}_D(X) < 0.05\,\max_X \mathcal{L}_D(X)$, while models with higher likelihood are shown with the blue color map. Darker shades indicate larger $\mathcal{L}_D(X)$; for readability, the color scale is obtained through a min-max normalization such that $0 \leq \mathcal{L}_D(X) \leq 1$, so that the best model's likelihood is 1. The colored lines represent the metamodel reconstructions in Sec.~\ref{sec_fits}: FSU2 lies outside of our informed prior because its nuclear matter parameters are outside the range compatible with the $\chi_{EFT}$ constraint discussed in Sec.~\ref{sec_par_space}.
    } \label{fig_mass-radius}
\end{figure}

The mass-radius relation is presented in Fig.~\ref{fig_mass-radius}, where each line $R_X(M)$ corresponds to a different instance of $X$. The gray background indicates the informed prior distribution, while the posteriors are shown in varying shades of blue representing the likelihood value. 
The informed prior spans radii from 9 to 15\,km and maximum masses $M_{\mathrm{TOV}}$ from $\sim 1.4$ to $\sim 3\,M_{\odot}$. These ranges illustrate the flexibility of our parametrization: although our prior coverage is narrower than that of composition-agnostic approaches\footnote{
    For example, barotropic EoS families built with Gaussian-process models can generate priors where radii span from $\sim 6$ to 16\,km \citep{Essick_2019,Landry_2020} and $M_{\mathrm{TOV}}$ up to $\sim 3.5\,M_{\odot}$~\citep{Mroczek_2023}.
}, we obtain remarkably similar mass-radius posteriors once the astrophysical filters are applied (i.e., after the full information $D$ is taken into account). 

In Fig.~\ref{fig_mass-radius} we also show the posterior of $M_{\mathrm{TOV}}$ against the informed prior, where one can see the combined effect of the filters. Radio-timing observations require $M_{\mathrm{TOV}}\gtrsim2\,M_{\odot}$, while a maximum mass above $2.6\,M_{\odot}$ is disfavored by the LVK constraint and the newest NICER measurements, which prefer a softer EoS.

\begin{figure}[htbp]
    \centering
    \includegraphics[width=\columnwidth]{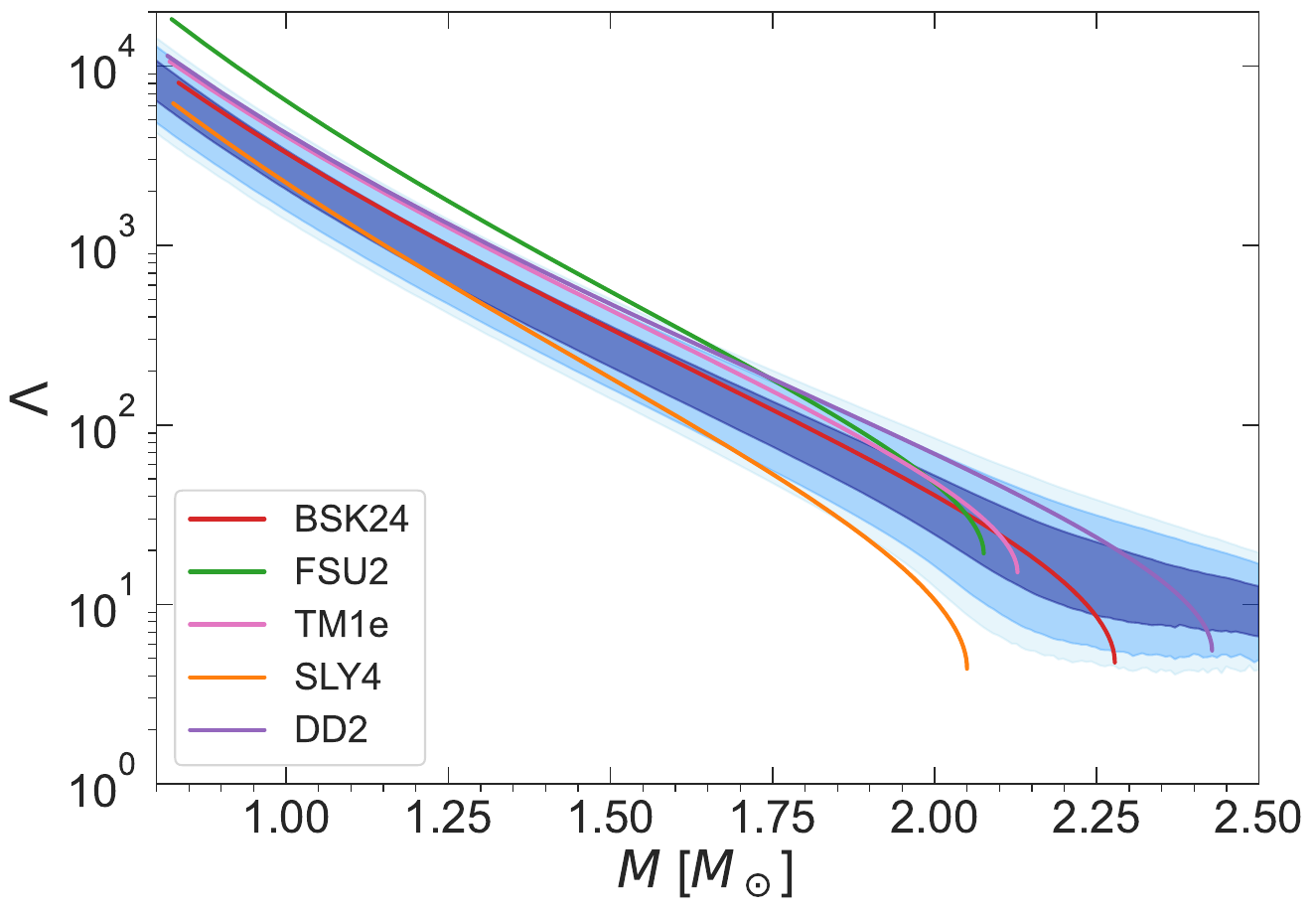}
    \caption{
    Posterior for the dimensionless tidal deformability $\Lambda_X$. The bands show the 68\%, 95\%, and 99\% quantiles for a given mass $M$. The colored lines denote the $\Lambda_X(M)$ relation for the metamodel fits.
    } \label{fig_lambda_m}
\end{figure}

Apart from $R_X(M)$, the mass-tidal deformability relation $\Lambda_X(M)$ is also largely composition-agnostic, making it a useful diagnostic for assessing whether the metamodel reproduces the overall trend and width of agnostic posteriors. This is shown in Fig.~\ref{fig_lambda_m}. The posterior displays the expected steep decrease with increasing $M$, reflecting the well-known dependence of $\Lambda_X$ on stellar compactness. To make the impact of the astrophysical and stability-causality filters more explicit, the slice at $M=1.4\,M_\odot$ is also shown in Fig.~\ref{fig_radditidal}, together with the prior.

The latest NICER data leave a clear imprint on the predicted radii and tidal deformabilities of a canonical NS, as shown in Fig.~\ref{fig_radditidal}. In the first two panels, which report the radius $R$ and tidal deformability $\Lambda$ of a $1.4\,M_\odot$ NS, the posterior shifts toward smaller values, as expected from a global softening of the EoS. In particular, when J0614 is included, the posterior median and 68\% quantiles for the radius move from $12.52^{+0.41}_{-0.42}$ to $12.21^{+0.39}_{-0.54}$, while those for the tidal deformability decrease from $506^{+130}_{-101}$ to~$435^{+97}_{-115}$.
Compared with previous agnostic \citep{Brandes_2025,Rutherford_2024} and nucleonic \citep{MalikSurvey} studies based on similar filters, we find slightly larger median values for both the radii and tidal deformabilities, even after the additional weighting toward softer EoSs induced by J0614. One possible source of this difference is the implementation of the pQCD constraint, which is not included in our framework and can itself soften the EoS~\citep{Koehn_2025,Somasundaram_2023}. Nevertheless, direct comparisons remain nontrivial because of differences in observational inputs and implementation choices. Within these limitations, our results with and without the newest pulsar remain compatible with the literature at the 68\% level, while the overall trend induced by the updated observational filter is qualitatively robust.

We then assess to what extent the same J0614-induced softening affects the expected f-mode frequencies. 
{To do this, we follow the strategy of \citet{Montefusco2025} (see equation (17) therein) to obtain synthetic f-mode frequencies based on the inversion of known quasi-universal relations valid beyond the Cowling approximation \citep{DebFullGR}.} This makes it possible to compute the f-mode frequencies for a large ensemble of models ($5\times10^5$) across the full NS mass range without solving the perturbation equations\footnote{
    { The calculation of synthetic frequencies is robust in the present context because the f-mode probes the $\beta$-equilibrated stellar sequence and depends essentially only on the equilibrated sound speed $v_\beta$. Thus, the use of quasi-universal relations is appropriate for this observable~\citep{Montefusco2025}. }
}. 
For a canonical $1.4\,M_\odot$ NS, the posterior shifts toward higher frequencies, again reflecting the fact that the new NICER data entering $\mathcal{L}_D(X)$ favor softer EoSs. 

\subsection{Threshold for dUrca processes}

\begin{figure*}[ht!]
    \centering

    \begin{minipage}[b]{0.4\textwidth}
        \centering
        \includegraphics[width=\textwidth]{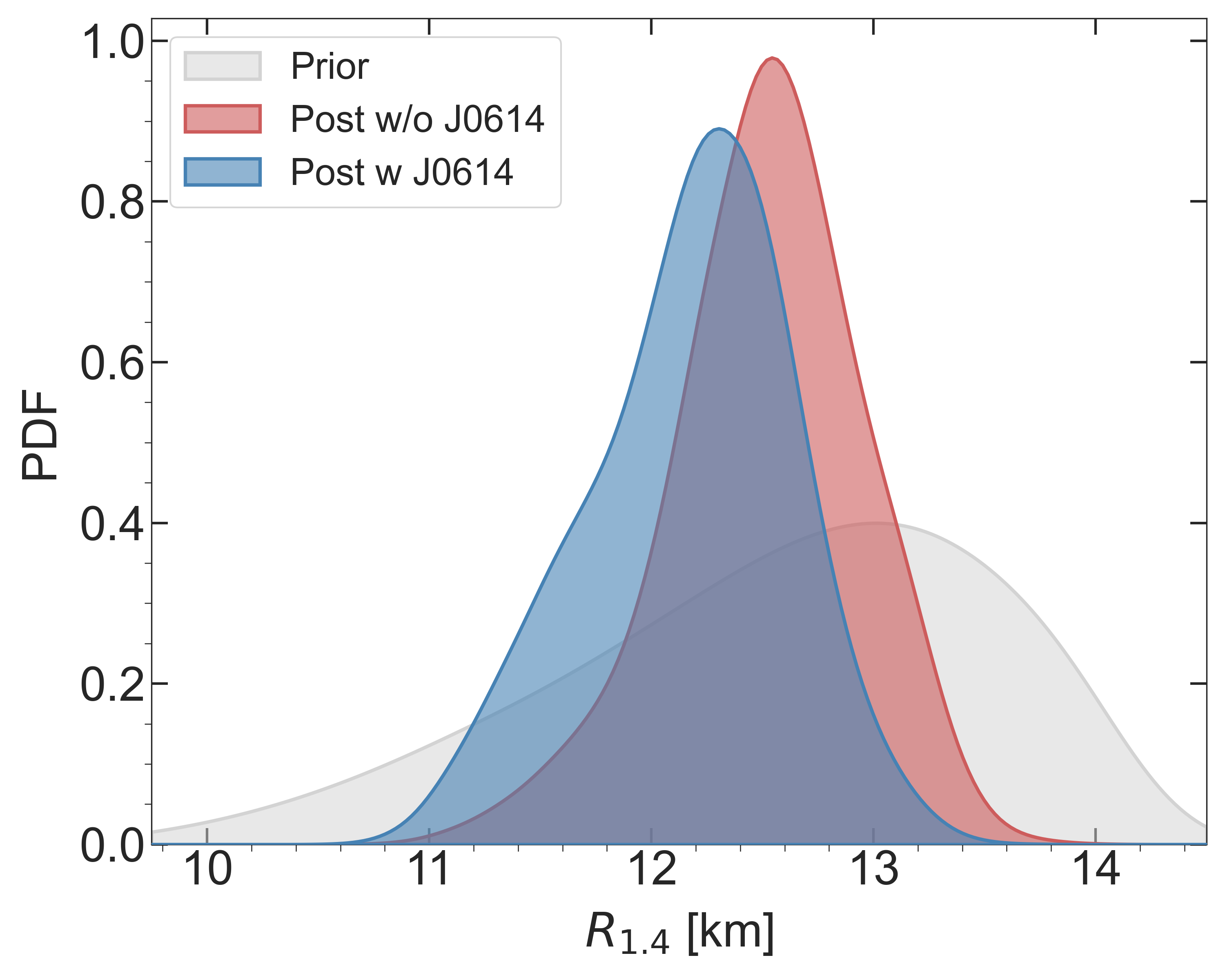}
    \end{minipage}
    \hfill
    \begin{minipage}[b]{0.4\textwidth}
        \centering
        \includegraphics[width=\textwidth]{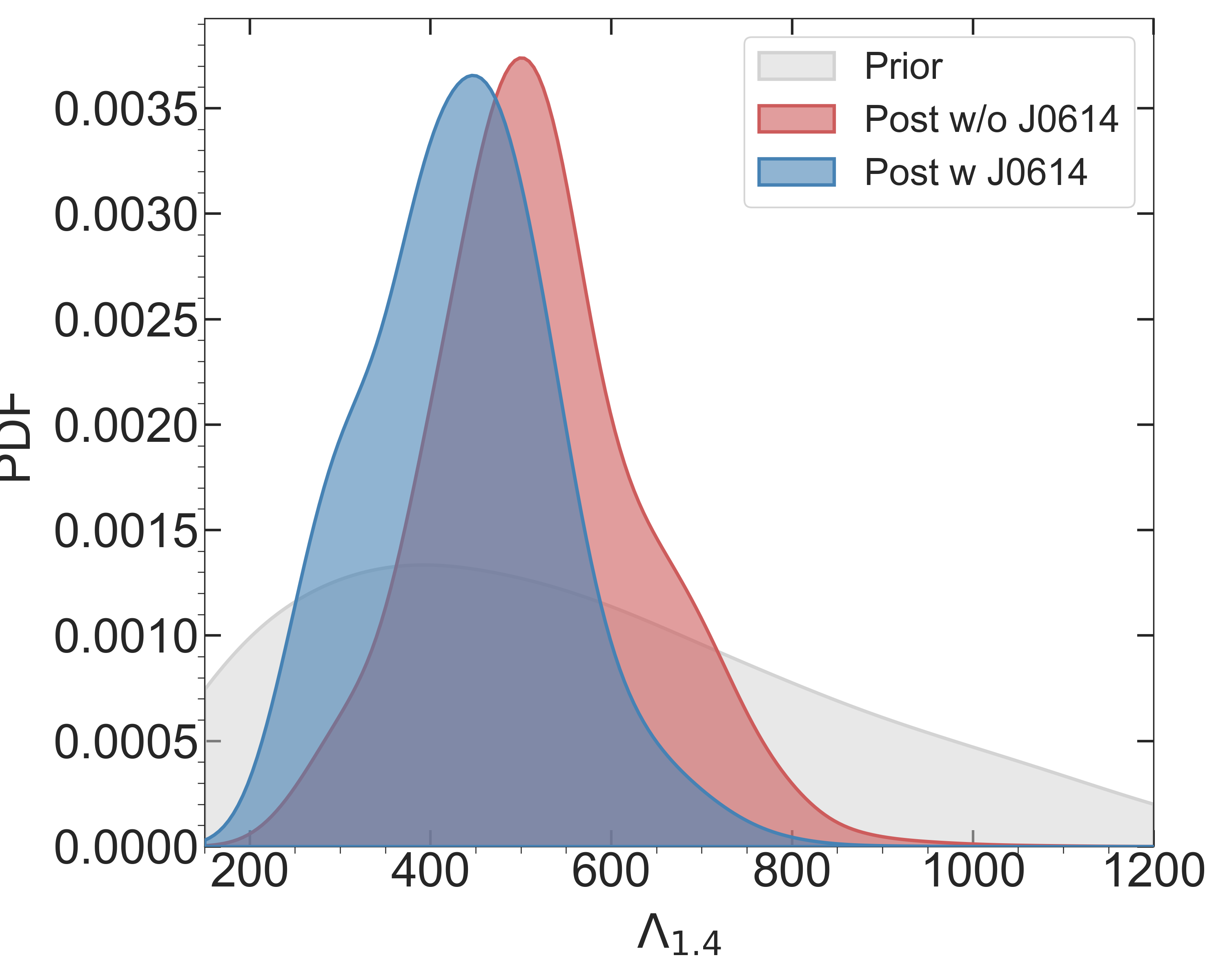}
    \end{minipage}
    
    \vskip\baselineskip

    \begin{minipage}[b]{0.4\textwidth}
        \centering
        \includegraphics[width=\textwidth]{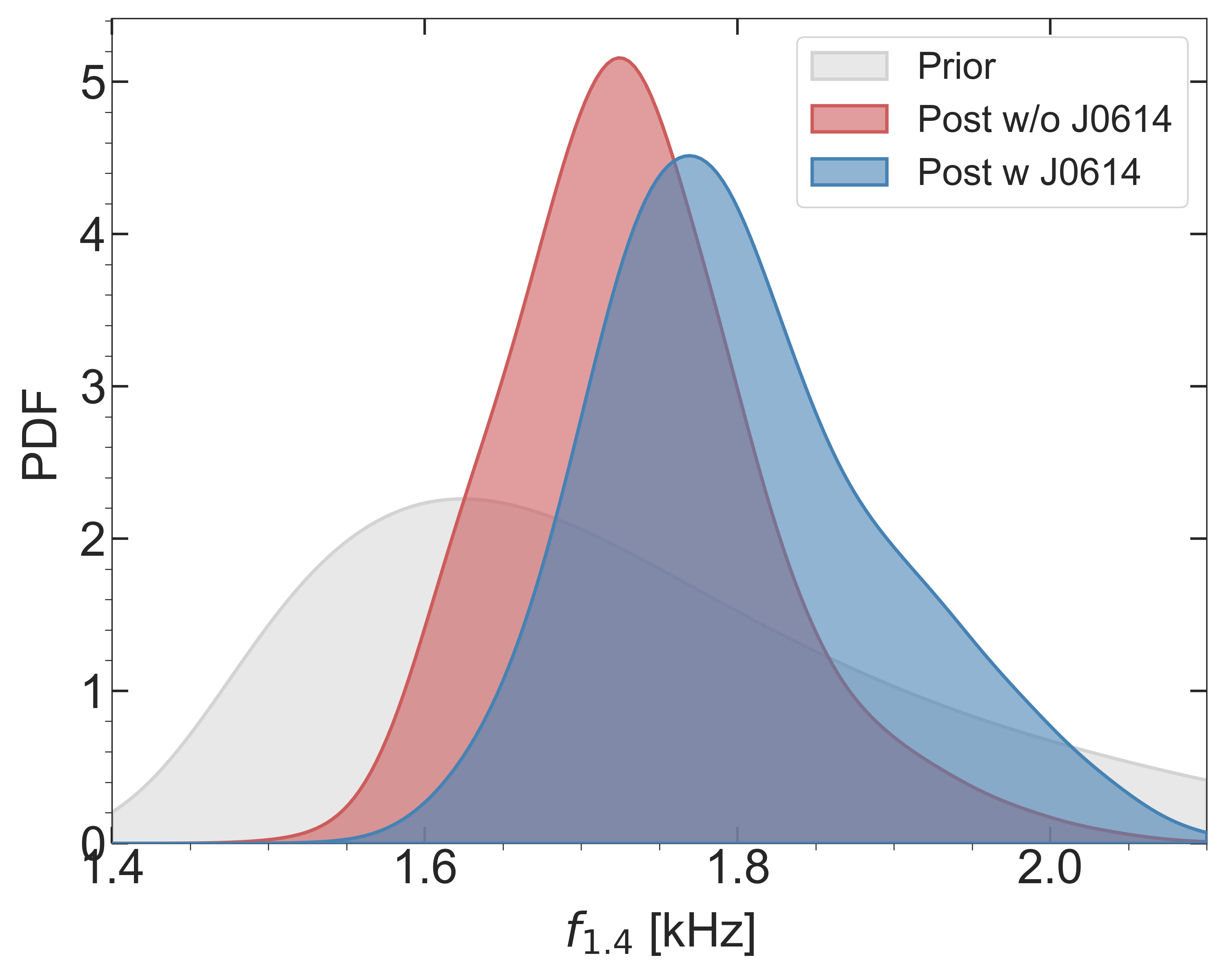}
    \end{minipage}
    \hfill
    \begin{minipage}[b]{0.4\textwidth}
        \centering
        \includegraphics[width=\textwidth]{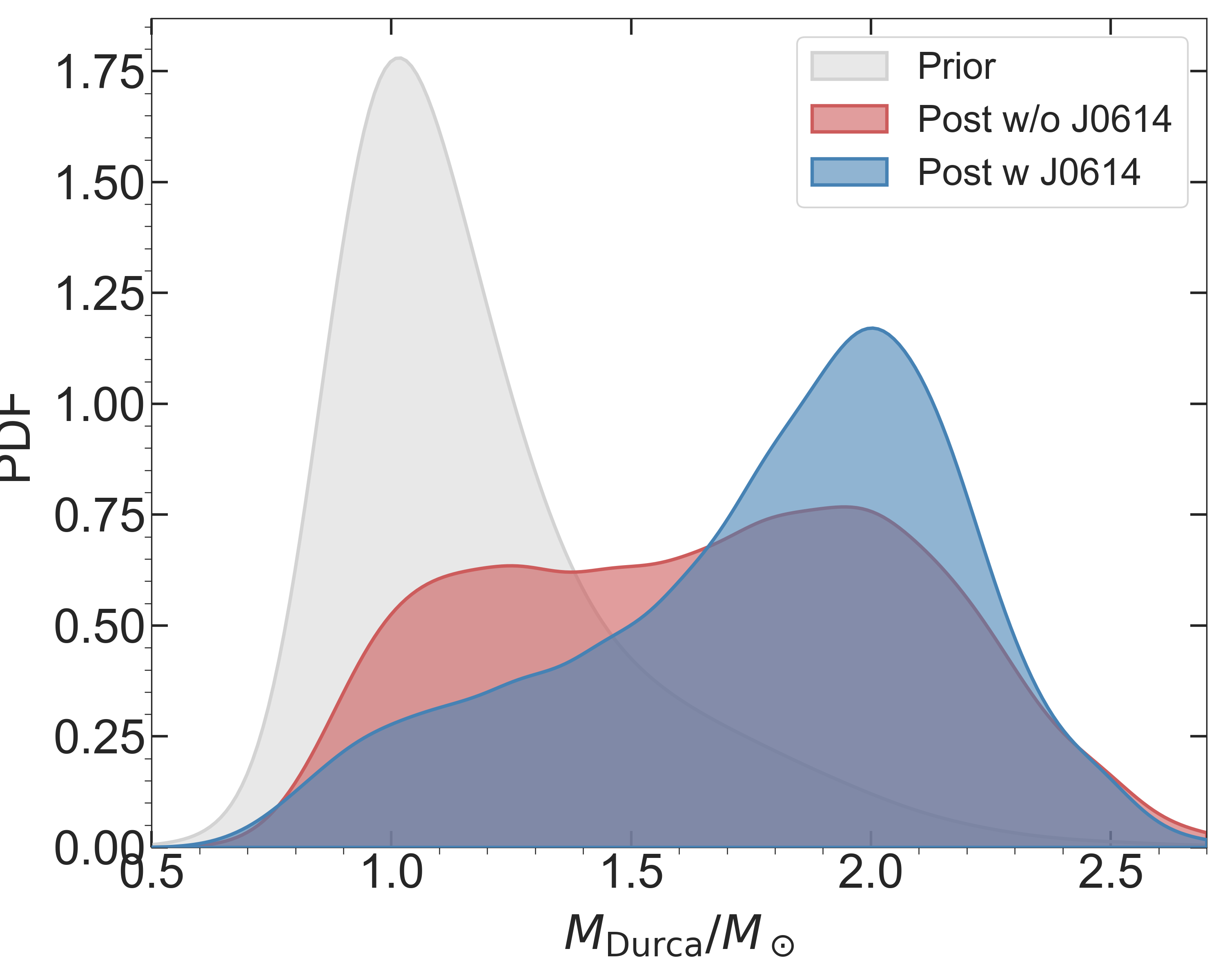}
    \end{minipage}

    \caption{
    Posterior distributions of the radius, tidal deformability, and f-mode frequency for a $1.4\,M_\odot$ neutron star with (blue) and without (red) the new NICER results. For comparison, the informed prior distributions are shown in gray. The last panel refers to the distribution of the dUrca threshold mass~$M_{\mathrm{dU}}$. 
    } \label{fig_radditidal}
\end{figure*}

We study the mass $M_{\mathrm{dU}}(X)$ at which the dUrca process\footnote{
    The muonic channel $n \rightarrow p+\mu+\bar{\nu}_\mu$ gives a more restrictive threshold and, therefore, would correspond to a larger~$M_{\mathrm{dU}}$. Hence, the electronic channel is taken as the relevant onset criterion for dUrca cooling.
} $n \rightarrow p+e+\bar{\nu}_e$ becomes kinetically allowed at the very center of an NS~\citep{lattimerUrca1991,sedrakian_urca_24}, assuming $\beta$-equilibrated $npe\mu$ composition.
To this end, we have to find the baryon number density $n^X_{\mathrm{dU}}$ that satisfies the implicit equation~\citep{klahn_PhysRevC_2006}
\begin{equation}
     x^\beta_n(n)^{1/3} =  x^\beta_p(n)^{1/3} + x^\beta_e(n)^{1/3} \, .
    \label{eq_DURCA}
\end{equation}
The threshold mass $M_{\mathrm{dU}}(X)$ corresponds to the mass of an NS with central density~$n^X_{\mathrm{dU}}$, see, e.g.,~\citep{MargueronMetaI,Malik_2022b,scurto2025delta}. We observe that, once the filters are applied, the posterior becomes almost flat, with a small peak that becomes more visible when the new J0614 results are included; see the last panel in Fig.~\ref{fig_radditidal}. 
This flattening is likely to be mostly driven by the $M_{\mathrm{TOV}}\gtrsim 2\,M_{\odot}$ requirement (point 3 in Sec.~\ref{sec_par_space}): since $M_{\mathrm{dU}}(X)<M_{\mathrm{TOV}}(X)$ is a hard boundary, imposing a lower limit on $M_{\mathrm{TOV}}(X)$ automatically reweights the sample toward models that can sustain larger values of $M_{\mathrm{dU}}(X)$. More generally, the astrophysical filters (including, to some extent, NICER data) suppress many of the models contributing to the low-mass peak of the informed prior and consequently enhance the relative weight of the high-$M_{\mathrm{dU}}$ tail. In this sense, the posterior appears to dominate the prior at large $M_{\mathrm{dU}}$ because of a redistribution of probability mass, while remaining strictly within the same support as the informed prior, even if this is not immediately apparent from the plot because of the very small prior tail at high masses.

\subsection{LVK posterior predictive}

\begin{figure*}
  \centering
  \subfigure{
    \includegraphics[width=0.48\textwidth]{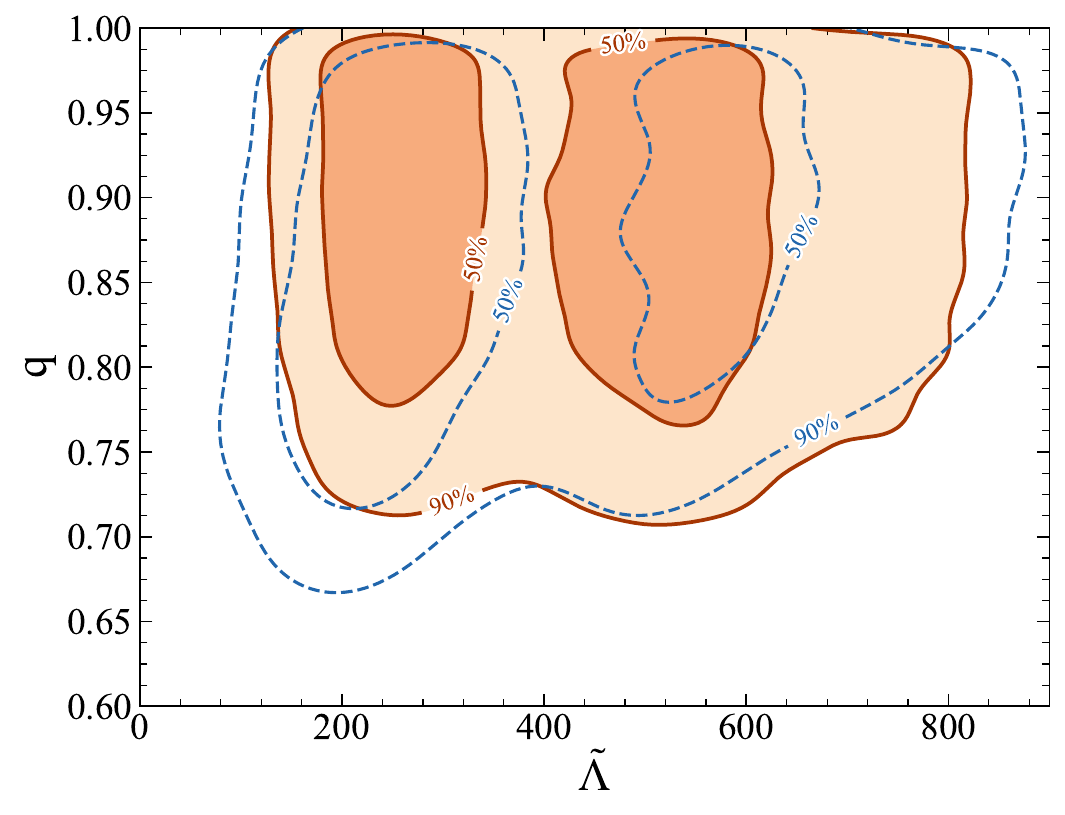}
  }
  \hfill
  \subfigure{
    \includegraphics[width=0.48\textwidth]{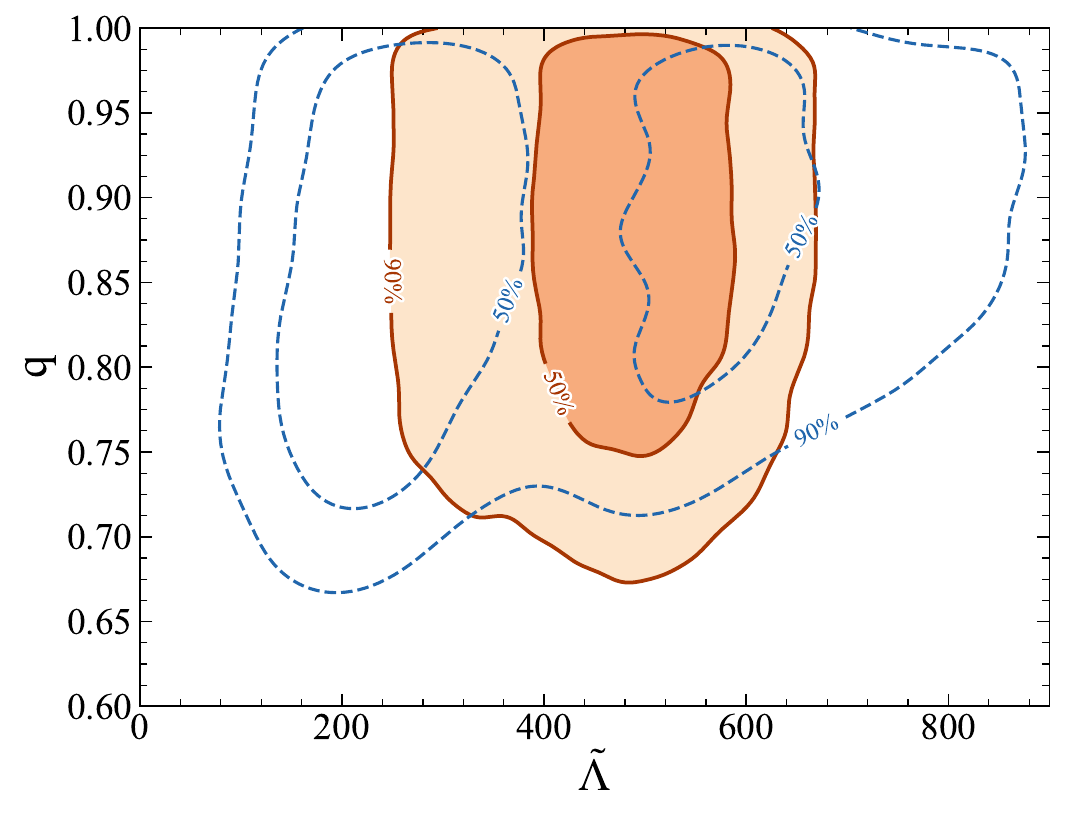}
  }
  \caption{
    Posterior predictive for GW170817 in the $(\tilde{\Lambda},\,q)$ plane, where $\tilde{\Lambda}$ is the effective tidal deformability of the binary system~\citep{GW170817_1} and $q = m_2/m_1 \leq 1$ is the mass ratio. Following the LVK convention~\citep{GW170817_1}, the filled orange contours show the 50\% and 90\% confidence regions obtained from our posterior predictive distribution. Left: posterior predictive derived solely from the informed prior plus the tidal likelihood, corresponding to point 5 in Sec.~\ref{sec_par_space}. Right: posterior predictive obtained with the full data $D$. Dashed blue contours show the 50\% and 90\% confidence regions of the LVK \texttt{IMRPhenomPv2\_NRT} low-spin posterior from GWTC-1~\citep{2019AbbottPRX}.
  }
  \label{fig:gw170817_predictive}
\end{figure*}

To clarify how our prior assumptions, namely the choice of a specific $e_X$ and its prior, together with the data $D$, shape the interpretation of the GW170817 tidal constraints, we show in Fig.~\ref{fig:gw170817_predictive} the posterior predictive distribution in the $(\tilde{\Lambda},q)$ plane, where $q=m_2/m_1\leq 1$ is the mass ratio and $\tilde{\Lambda}$ is the effective tidal deformability of the binary system. This distribution is obtained by drawing masses from the LVK posterior and calculating the corresponding tidal deformabilities from the weighted ensemble of our EoS models; see Sec.~\ref{sec_par_space} and App.~B of~\citep{Montefusco2025} for technical details.

The left panel of Fig.~\ref{fig:gw170817_predictive} shows that our posterior predictive already reproduces the LVK GW170817 contours rather well, despite the substantial prior structure entering our construction. The residual discrepancy with respect to the LVK result of \citep{2019AbbottPRX} appears at low values of $\tilde{\Lambda}\lesssim 300$, and becomes more pronounced as $q$ decreases. This effect is not naturally interpreted as the consequence of any single nuclear ingredient taken in isolation, but rather as the direct result of imposing a single common EoS for the two stars. Once the masses are drawn, the two tidal deformabilities are no longer free, but are linked by the same relation $\Lambda_X(M)$. At fixed chirp mass, this becomes especially restrictive at low $q$, where very small values of $\tilde{\Lambda}(m_1,m_2,\Lambda_X(m_1),\Lambda_X(m_2))$ are difficult to realize within a common-EoS construction. This is the main methodological difference with respect to \citep{2019AbbottPRX}, where the tidal sector was not constrained by a common EoS relation shared by the two objects.

The same mechanism was identified independently by \citet{magnall_2025ApJ}, and is even more evident in the right panel of Fig.~\ref{fig:gw170817_predictive}. Using physics-informed priors that tie masses and tidal deformabilities through the EoS in their parameter-estimation procedure for GW170817, they found a strong suppression of the low-$\tilde{\Lambda}$ tail, obtaining the 90\% interval $226 \le \tilde{\Lambda}_{1.186} \le 690$. Our right panel, where the full data set $D$ of Sec.~\ref{sec_par_space} is used, shows the same effect, with a 90\% lower cutoff around $\tilde{\Lambda}\sim 250$, in very good agreement with their result.

The comparison between the two panels of Fig.~\ref{fig:gw170817_predictive} therefore shows that the main origin of the discrepancy with respect to the original LVK posterior is not a peculiarity of our metamodel, but the generic effect of imposing common-EoS consistency between the two objects of GW170817. In this sense, the posterior predictive distribution provides a useful diagnostic of how much of the tidal inference is driven by the observational likelihood itself and how much by the physical prior structure relating masses and tidal deformabilities.

\subsection{Microscopic properties: speed of sound and composition}

\begin{figure}
    \centering
    \includegraphics[width=\columnwidth]{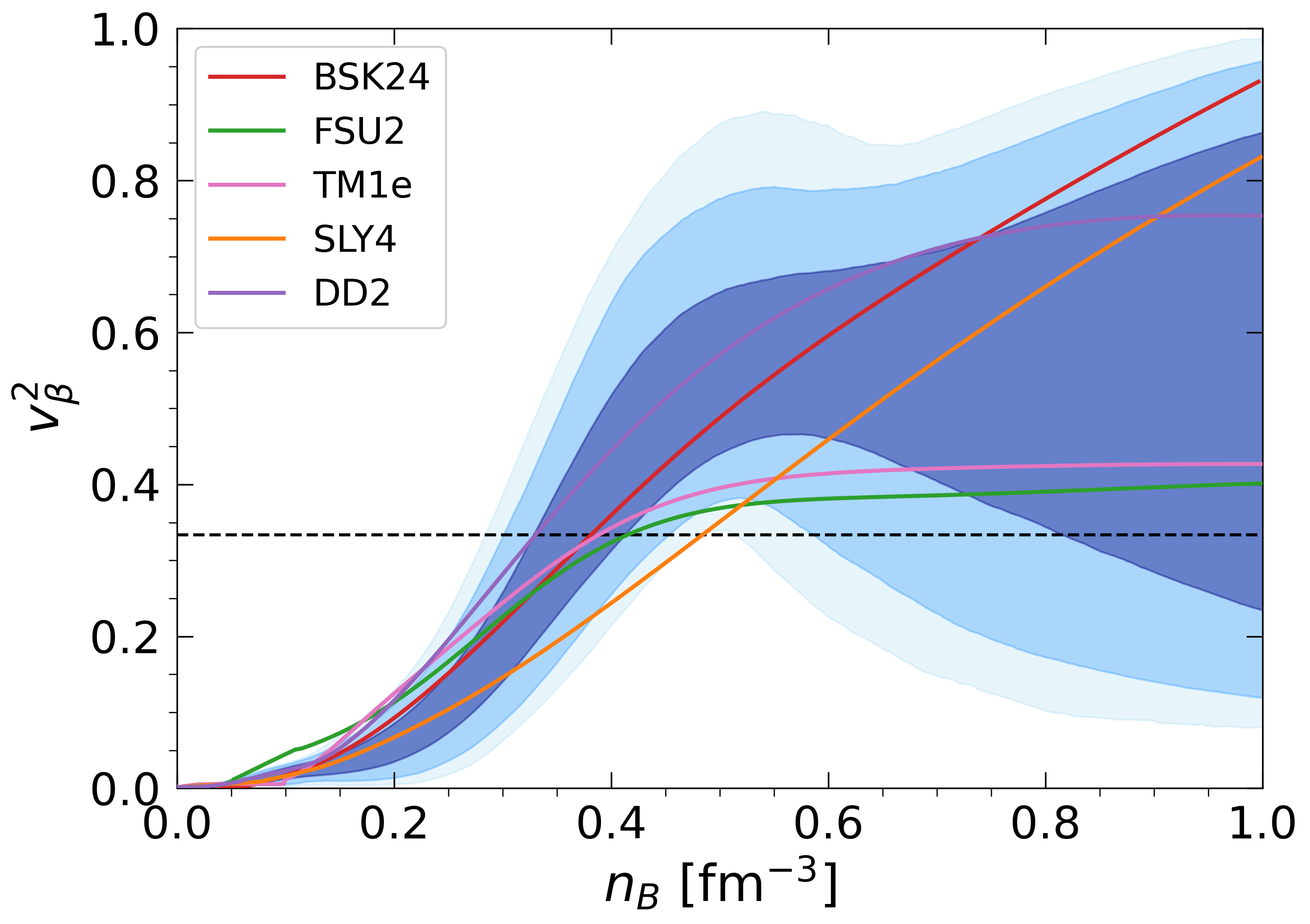}
    \caption{Posterior of the equilibrated squared sound speed $v^2_{\beta X}(n)$. The bands show the 68\%, 95\%, and 99\% quantiles at fixed baryon number density~$n$. As in Fig.~\ref{fig_mass-radius}, the colored lines denote the metamodel fits. The horizontal dotted line indicates the benchmark value~$v^2_{\beta}=1/3$.
    } \label{fig_speedofsound}
\end{figure}

The squared sound speed $v^2_{\beta X}$ tends to exhibit a monotonic increase for most of the sampled metamodel instances, as shown in Fig.~\ref{fig_speedofsound}. This is consistent with previous studies based on RMF (e.g., \citep{char_metaRMF_2023,Malik_2024,char_metaRMF_2025}) and agnostic sound speed barotropic models~\citep{Brandes_sound_inference_23,Brandes_2023,Brandes_2025}. In particular, our posterior in Fig.~\ref{fig_speedofsound} closely resembles the agnostic posterior found by \citet{Brandes_2025}, except that, at high densities, our distribution extends slightly farther below the conformal limit (i.e., $v^2_\beta=1/3$) within the 68\% credible region.

Above $\sim 0.5\,$fm$^{-3}$, we observe an increase in the spread of the posterior distribution of $v^2_\beta$, associated with a decrease in the lower quantiles. This behavior, as well as visual inspection of several curves $v^2_{\beta X}(n)$, suggests that a substantial fraction of our models exhibits a non-monotonic speed of sound, a possibility not excluded by current agnostic studies~\citep{altiparmak2022ApJ,Brandes_2025}. This results in a broad peak in $v^2_{\beta X}(n)$ for some of the sampled metamodel instances. Although features in the sound speed can be linked to properties of the EoS \citep{Mroczek2024PhRvD}, we stress that the possible presence of a smooth bump in $v^2_{\beta X}(n)$ is not associated with any particular physical feature of our~$\epsilon^{tot}_X$.\footnote{
    By contrast, the appearance of muons around $\sim(0.6--0.8)n_0$ produces a nonanalytic feature in $v^2_{\beta X}(n)$ at the density where $x_\mu^\beta$ first becomes nonzero. At even lower densities, around $\sim0.5n_0$, the crust-core transition may also induce a nonanalytic feature in $v^2_{\beta X}(n)$. These physical features are, however, quite small and not visible in Fig.~\ref{fig_speedofsound}.
} 
The fact that our posterior contains instances with a non-monotonic $v^2_{\beta X}$ is a consequence of both our prior, namely that \eqref{eq_u02} does not enforce monotonicity of the sound speed, and the causal filtering imposed on the ensemble: since $v^2_{\beta X}$ cannot exceed unity, models that rise rapidly at intermediate densities must eventually flatten or decrease at higher densities. 

One advantage of using a composition-aware metamodel is that it allows one to explore deviations from $\beta$ equilibrium~\citep{Montefusco2025}. A simple application is the assessment of the convective stability of NS cores~\citep{Reisenegger_1992,Friedman_2017}, which is linked to the positivity of
\begin{equation}
    \label{eq_vsdiff}
    \Delta = \frac{1}{v^2_{\beta}} - \frac{1}{v^2_{f}} \geq 0 \, ,
\end{equation}
where $v_{\beta}$ and $v_{f}$ are the equilibrated and frozen sound speeds, respectively. 
The inequality \eqref{eq_vsdiff} is essentially a Ledoux criterion for a stratified fluid and ensures that displaced fluid elements experience a restoring force, implying stability against convection~\citep{Reisenegger_1992,Lai1994,Jaikumar21}.  
{Indeed, the difference between the inverse squared sound speeds, $\Delta$, is directly related to the Brunt-V{\"a}is{\"a}l{\"a} frequency, $N$, which is particularly relevant for g-modes:}
\begin{equation}
    N^2(r) = g^2(r) \Delta(r) e^{\nu(r)-\lambda(r)},
\end{equation}
where $g=(dP_\beta/dr)(\epsilon_\beta+P_\beta)^{-1}$, and $\nu$ and $\lambda$ are the metric functions; all quantities are evaluated along the usual Schwarzschild radial coordinate $r$~\citep{Dommes_2016}.
The sign of $\Delta$ determines the sign of $N^2$, and hence whether the star is convectively stable ($N^2 > 0$) or unstable ($N^2 < 0$)~\citep{Lai1994,Jaikumar21,Tran2023,ZhaoLattimer2022}. 

\begin{figure}[htbp]
    \centering
    \includegraphics[width=\columnwidth]{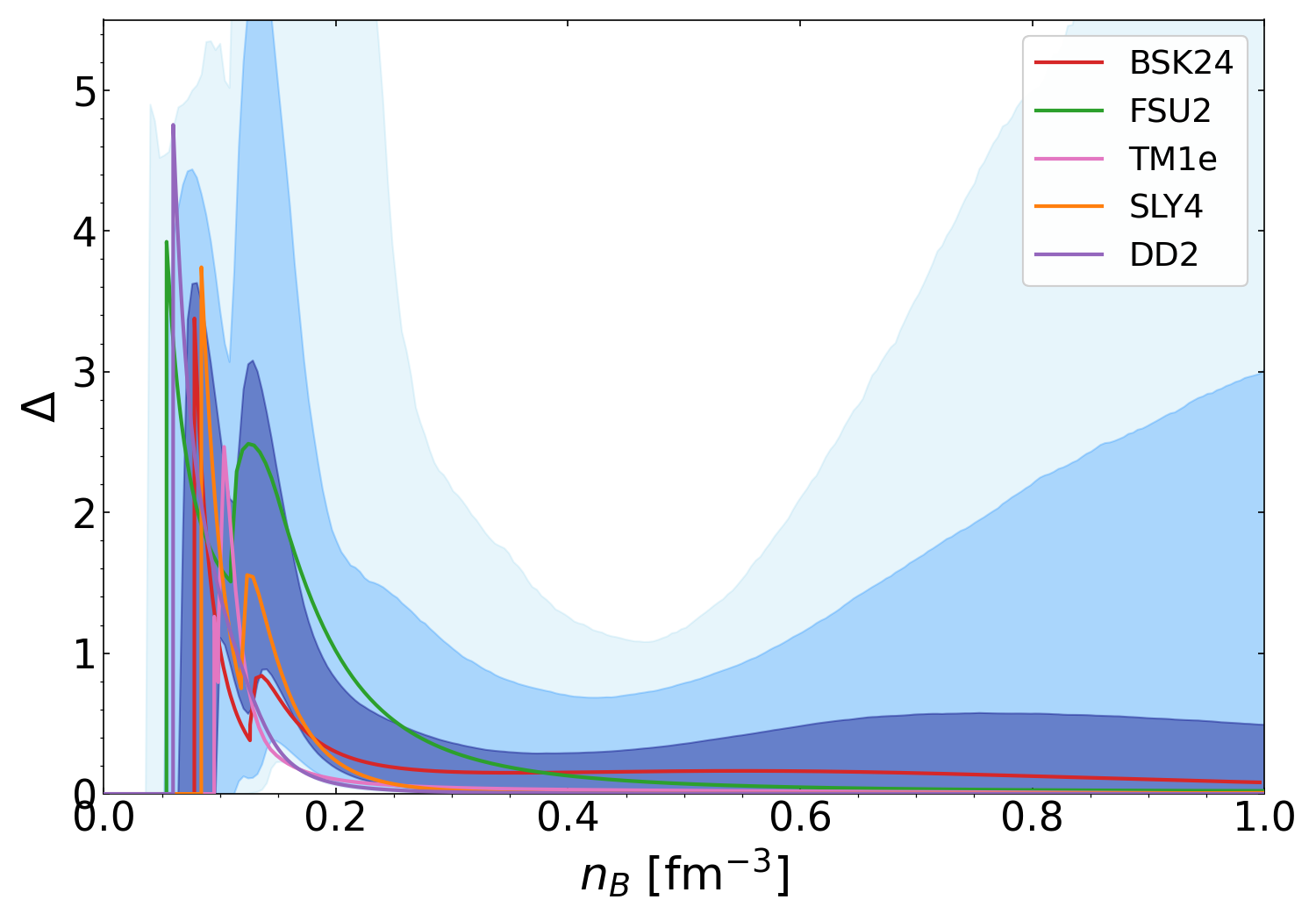}
    \caption{Posterior of $\Delta_X=1/v^2_{\beta X}-1/v^2_{f X}$, following the same color scheme as Figs.~\ref{fig_mass-radius} and~\ref{fig_speedofsound}.
    } \label{fig_vsdiff}
\end{figure}

Figure~\ref{fig_vsdiff} shows the posterior of $\Delta_X=1/v^2_{\beta X}-1/v^2_{f X}$ as a function of baryon density. 
{Two distinct features are visible. The first, located near the crust-core transition, is not physically meaningful in the present context: because the crust is solid, the simple homogeneous-fluid Ledoux criterion is not reliable there, and we therefore set $\Delta_X=0$ in the solid crust. The second feature is physical and corresponds to the onset of muons; it reflects the fact that, while $v_{f X}(n)$ remains continuous across the muon threshold, $v_{\beta X}(n)$ develops a nonanalytic feature because of the change in chemical composition.}

{
Moreover, Fig.~\ref{fig_vsdiff} shows that $\Delta_X>0$ throughout the homogeneous core, up to and beyond $1\,\text{fm}^{-3}$. 
Indeed, the filter $0<v_{\beta X}^2<v_{f X}^2<1$ automatically implies that the Ledoux criterion is satisfied throughout the core: since~\eqref{eq_vsdiff} is satisfied across all relevant densities, all the NSs in our posterior can support stable g-modes~\citep{Reisenegger_1992}.
}

\begin{figure}[htbp]
    \centering
    \includegraphics[width=\columnwidth]{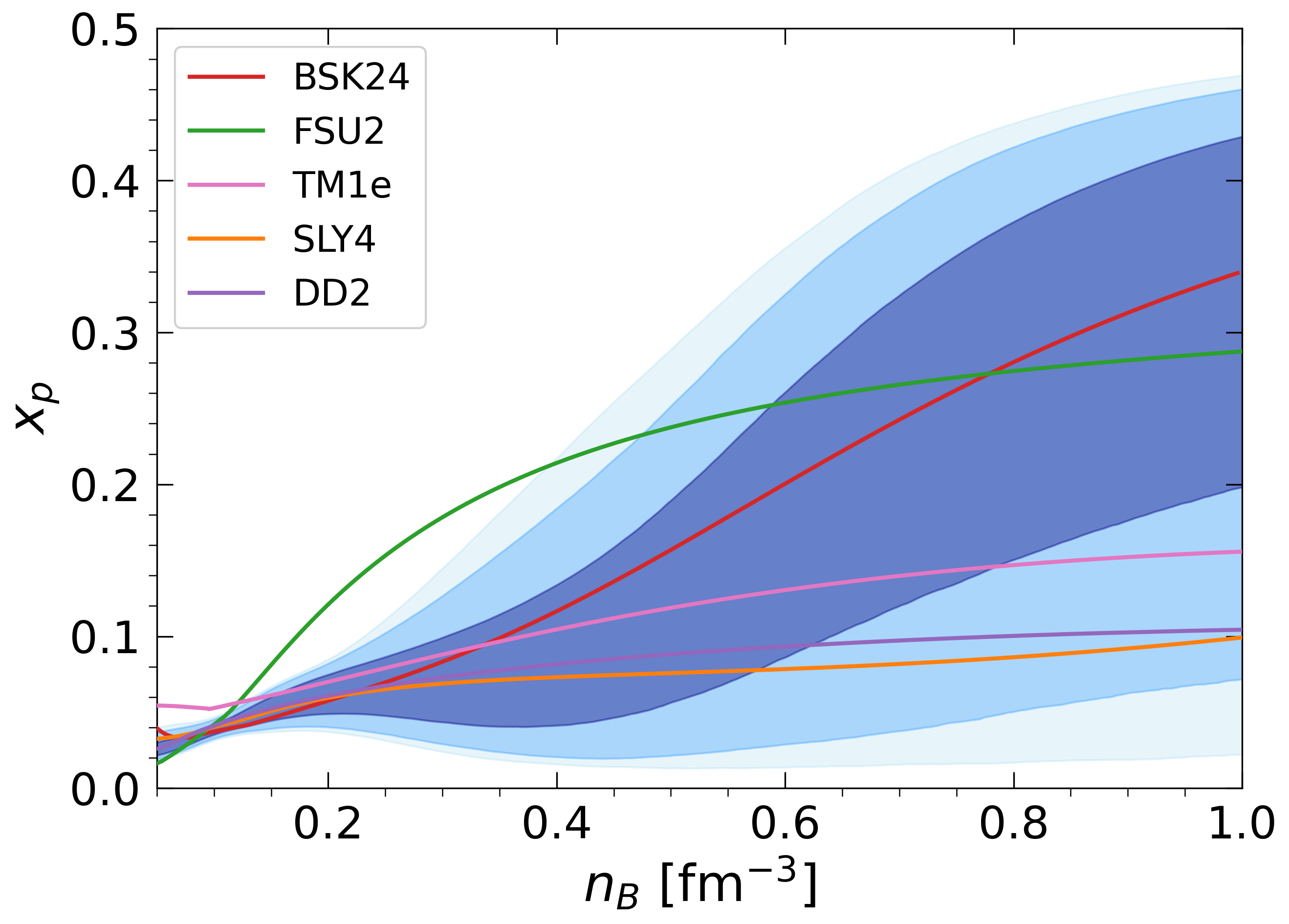}
    \caption{
    Posterior for the proton fraction $x^\beta_p$ as a function of baryon density. The shaded bands represent the corresponding credible regions, following the same color scheme as Figs.~\ref{fig_mass-radius} and~\ref{fig_speedofsound}. 
    } \label{fig_composition}
\end{figure}

Another interesting quantity is the internal composition of NS matter, particularly the proton fraction $x^\beta_p$, shown in Fig.~\ref{fig_composition} as a function of baryon density. We observe a general increase of $x^\beta_p(n)$, with all models within the 68\% credible interval exceeding $\sim0.2$ at $n\sim 1\,$fm$^{-3}$. As for the sound speed, the proton fraction exhibits a broad posterior, indicating that the current set of constraints implemented in $D$ does not strongly restrict its behavior. The fact that essentially the whole range $\sim0$--$0.5$ is spanned at high densities, where there is no information in $D$ that can directly constrain composition, provides a measure of the flexibility of our $u_i$ in~\eqref{eq_u02}.

{
Fig.~\ref{fig_composition} can be compared with previous studies based on composition-aware metamodeling schemes, where the asymptotic causality requirement was addressed using RMF Lagrangians with parametrized density-dependent couplings; see Fig.~7 in \citep{char_metaRMF_2023} and Fig.~4 in \citep{char_metaRMF_2023,scurto2025delta}. Although the data sets, likelihood implementations, and prior assumptions are not exactly identical, the comparison is informative because all these analyses combine very similar low-density nuclear information with present astrophysical constraints under a nucleonic hypothesis.}
{
The broader posteriors obtained in this work should therefore be interpreted as a consequence of the larger functional freedom allowed by the metamodel. Since present observations and nuclear-theory constraints do not strongly determine the composition at suprasaturation density, narrow high-density composition posteriors typically reflect correlations imposed by the chosen functional class. Broader posteriors, by contrast, provide a more conservative propagation of the current lack of information, which is one of the purposes of the present metamodel}\footnote{
    { The precise posterior widths remain, of course, conditional on the adopted metamodel scheme. Future comparisons with other composition-aware parametrizations will therefore be useful to quantify residual model dependence. This is a general feature of EoS inference rather than of the present construction: fully agnostic representations, including nonparametric ones, can also yield different posterior structures when different priors or parametrizations are used~\citep{Rutherford_2024,biswasPRD2026}. }
}. 

{
This is particularly relevant for the lower tail of the proton fraction at high density. Allowing the prior to explore configurations with $x_p<0.1$ in the core is important for an unbiased estimate of the possible onset of dUrca processes in NS cooling, as discussed above.}

\subsection{Nuclear matter parameters}

A convenient feature of the metamodel is that six of its parameters, namely $h_i^{(k)}$ with $k=0,1,2$ and $i=0,2$, can be mapped exactly onto the six standard NMPs $\{n_0,E_0,K_0,E_2,L_2,K_2\}$, see App.~\ref{app_mapping}. Since some prior information on the NMPs is available, these quantities are sampled directly in our Bayesian analysis using flat priors over reasonable intervals, reported in Tab.~\ref{tab_prior-bounds}. 
By allowing ranges broader than those suggested by nuclear experiments alone~\citep{rocamaza2018PrPNP}, we incorporate the available nuclear-data information conservatively and with minimal bias. This treatment could be extended in future work by adding to the data set $D$ constraints from posterior distributions obtained in Bayesian analyses of nuclear data~\citep{Klausner_2024}.
Although flat, this prior is informative; however, it contains no structure beyond the support of the chosen intervals and, in particular, no correlations among the NMPs. It is therefore interesting to check whether correlations among the NMPs emerge in the posterior.

\begin{figure*}
    \centering
    \includegraphics[width=2\columnwidth]{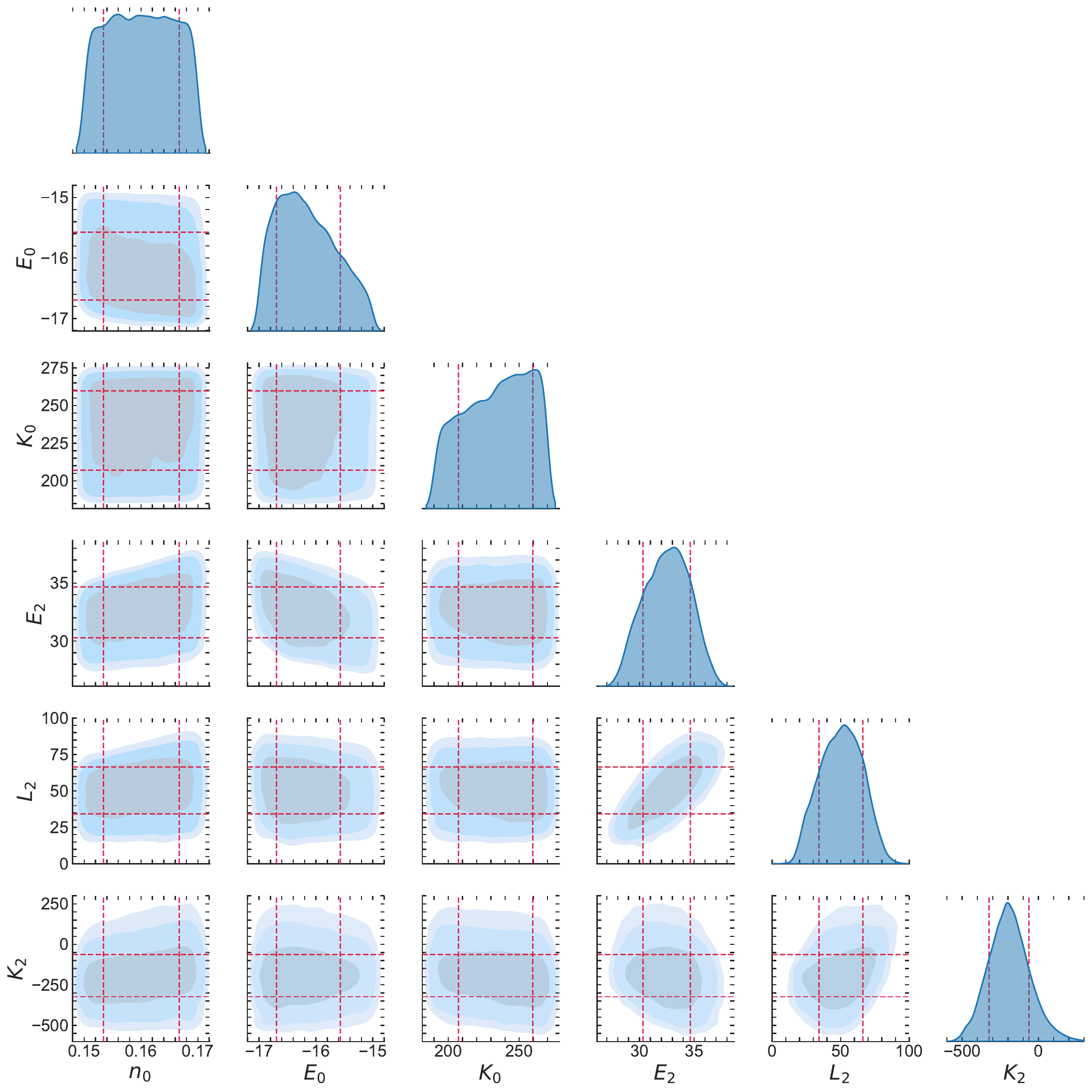}
    \caption{Corner plot of the posterior distribution of the NMPs $\{n_0,E_0,K_0,E_2,L_2,K_2\}$. The shaded regions denote the 68\%, 95\%, and 99\% credible regions. The parameters associated with $u_0$ remain largely unconstrained, except for $E_0$, whereas those associated with $u_2$ are shaped by the $\chi_{EFT}$ filter and by the maximum-mass constraint.}
    \label{fig_params}
\end{figure*}

The corner plot of the posterior distribution of the NMPs is shown in Fig.~\ref{fig_params}. The first three rows correspond to the parameters associated with $u_0$. We find that $n_0$ and $K_0$ remain essentially unconstrained, while $E_0$ is constrained by the fit to the AME2020 mass table.

By contrast, the parameters associated with $u_2$ are more affected by the data entering the likelihood. The posteriors of $E_2$ and $L_2$ are constrained primarily by the $\chi_{EFT}$ filter, which also induces small correlations with $K_2$. The latter is further constrained by the requirement that the models support at least the mass of PSR J0740+6620. Through its correlation with $K_2$, this requirement also induces a mild effect on the posterior of $L_2$.

\begin{table}[ht]
\centering
\begin{tabular*}{\columnwidth}{@{\extracolsep{\fill}}lccccccc@{}}
\toprule
& \textbf{Units} & \textbf{Median} & \multicolumn{2}{c}{\textbf{68\% CI}} & \multicolumn{2}{c}{\textbf{95\% CI}} \\
 &  &  & \textbf{Min} & \textbf{Max} & \textbf{Min} & \textbf{Max} \\
\midrule
$n_0$   & fm$^{-3}$ & 0.160  & 0.153  & 0.167   & 0.151   & 0.169 \\
$E_0$   & MeV       & -16.2  & -16.7  & -15.6   & -16.9   & -15.1  \\
$K_0$   & MeV       & 236    & 207    & 260     & 193     & 268    \\
$E_2$   & MeV       & 32.6   & 30.3   & 34.7    & 28.7    & 36.2   \\
$L_2$   & MeV       & 50.9   & 33.3   & 66.4    & 22.5    & 78.2   \\
$K_2$   & MeV       & -195   & -323   & -60     & -443    & 104    \\
\bottomrule
\end{tabular*}
\caption{
    Posterior medians and 68\% and 95\% credible intervals for the nuclear matter parameters.
}
\end{table}

\section{Conclusions}

We have presented an asymptotically causal nucleonic metamodel that preserves the main practical strengths of the original \citet{MargueronMetaI} scheme, namely its analytic structure, low computational cost, and exact mapping to the standard NMPs at saturation, while improving its high-density behavior. In particular, the proposed $e_X$ yields a better-controlled supranuclear regime and reduces the occurrence of pathological models with superluminal sound speeds or mechanical instabilities found in previous metamodel-based inferences~\citep{MargueronMetaII,HoaUniverse,davis2024,Montefusco2025}. 
Despite adopting the more restrictive stability-causality requirement $0 < v^2_\beta < v^2_f < 1$ as a hard filter to select metamodel instances consistent with special relativity and chemical stability, see~\citep{camelio_I,Montefusco2025}, the new asymptotically causal form retains $\sim 30$--$50\%$ of the sampled instances, compared with the $\sim 1$--$10\%$ typical of the previous implementation, depending on the prior $P(X)$ and on the precise filters and data $D$ entering $\mathcal{L}_D(X)$. This substantially increases the number of significant posterior samples and improves the statistical robustness of the inference.

The revised metamodel is flexible enough to reproduce, with good accuracy, representative EoSs from both Skyrme-like and RMF-like families. Although the fit is constrained only on the SNM and PNM slices, the reconstructed two-dimensional energy landscape $e_X(n,\delta)$ also reproduces the corresponding $\beta$-equilibrated EoS at the few-percent level for the pressure, energy density, composition, and sound speed in the core-density range. This is a nontrivial test for NS applications, since it probes the interpolation in isospin asymmetry of the proposed parametrization $e_X(n,\delta)$. In principle, fits of $e_X(n,\delta)$ to realistic EoSs may therefore be used as analytic representations of those EoSs in place of two-dimensional tables.

At the inference level, the main practical gain is that the asymptotically causal formulation makes Bayesian sampling of composition-aware EoSs much more efficient, even under the restrictive condition $0 < v^2_\beta < v^2_f < 1$, which also automatically guarantees that the Ledoux criterion for the convective stability of the NS core is satisfied~\citep{Reisenegger_1992,Lai1994}. Hence, our posterior is not only consistent with current mass, radius, and tidal deformability constraints, but also contains only stellar 
configurations that are convectively stable and can support g-modes~\citep{Reisenegger_1992}, while also satisfying the causality-stability condition for chemical mixtures~\citep{camelio_I}.

The use of composition-aware metamodels allows one to explore quantities that are inaccessible to composition-agnostic barotropic parametrizations, such as the proton fraction, the dUrca threshold, the Ledoux discriminant, or crustal properties. 
The present framework therefore provides a practical tool for statistical studies of NS physics beyond the purely barotropic sector. It also offers a computationally inexpensive complement to RMF-based metamodel studies~\citep{ScurtoPrediction,char_metaRMF_2023,char_metaRMF_2025,scurto2025delta,MalikSurvey}. 

At the same time, the analysis shows where current data cease to be constraining beyond the barotropic sector: while global stellar observables are significantly constrained, microscopic quantities in the inner core remain broadly distributed. 
{ This is not a defect, but an indication of the degree of agnosticism of the proposed metamodel; specifically, it reflects the limited constraining power of present static observables beyond the $\beta$-equilibrated barotropic sector \citep{mondal2022}. The framework will become more constraining only when genuinely composition-sensitive observables, such as cooling constraints, transport-sensitive signatures, or robust asteroseismic measurements, are available (see, e.g.,~\cite{montoli2020A&A}). }

There are also clear directions for refinement. First, the present EoS reconstructions are not very accurate below $\sim 0.5 n_0$, which is also the density regime relevant for the crust. Improving the low-density sector of the nucleonic metamodel has already been achieved in \citep{burrello2025crust} through the implementation of a universal low-density expansion for nuclear matter, and the same strategy can, in principle, be adopted here. Second, the recent emulator-assisted inferences of the NMPs by \citet{Klausner_2024}, obtained from a detailed match to nuclear masses and other nuclear properties, could be incorporated as a structured informative prior, as done in~\citep{klausner2025prc}. Both improvements would increase the reliability of the scheme in the crustal layers and enable more robust inferences of crust properties.

Finally, we stress that the specific parametrization $e_X$ proposed here is only one among infinitely many possibilities, and even simpler or better realizations may well exist. 
As with composition-agnostic schemes, finding a good metamodel scheme is therefore a matter of trial and error followed by a validation procedure to assess its flexibility and coverage of the space of all EoSs consistent with current knowledge. Within the present analysis, we find no obvious pathology that can be traced to the chosen implementation of $e_X$ or to the prior $P(X)$, but this is precisely why alternative metamodel schemes (e.g., \citep{MargueronMetaI,Huth2021,Lim2024,char_metaRMF_2023,ScurtoPrediction}) and purely agnostic schemes remain useful as validation and cross-checking tools.


\begin{acknowledgments}
We thank Philip John Davis, Anthea Fantina, Pietro Klausner, Luigi Scurto, Hoa Dinh Thi, and Stefano Burrello for interesting comments and feedback.
Partial support comes from the IN2P3 Master Project ``Mod\'elisation des Astres Compacts'' (MAC), the ANR project ``Gravitational waves from hot neutron stars and properties of ultra-dense matter'' (ANR-22-CE31-0001-01), and the CNRS International Research Project (IRP) ``Origine des \'el\'ements lourds dans l'univers: Astres Compacts et Nucl\'eosynth\`ese''~(ACNu).
\end{acknowledgments}


\appendix

\onecolumngrid

\section{Mapping to the nuclear matter parameters}
\label{app_mapping}

We give the exact algebraic map from the coefficients in the rational ansatz for $u_i(n)$ to the chosen NMPs at saturation.
For $i=0,2$, the six parameters $h_i^{(0,1,2)}$ in \eqref{eq_u02} are related to the empirical NMPs by matching the expansion of the energy per baryon $e_X(n,\delta)$ in \eqref{eq_eX} around saturation density to the reference phenomenological expression in \eqref{eq_snm}.
This ensures that the model $e_X(n,\delta)$ exactly reproduces, by construction, the chosen NMPs $\chi_0=\{E_0,L_0=0,K_0\}$ and $\chi_2=\{E_2,L_2,K_2\}$, as well as a chosen value of the saturation density $n_0$.
The two remaining parameters $h_i^{(3)}$ are free and are independently sampled at the prior level, while the procedure for sampling the prior of $a_i$, $b_i$, and $c_i$ is described in App.~\ref{app_abc}.

Since the mapping between the $h_i^{(0,1,2)}$ and the NMPs involves derivatives of the kinetic term $e_F(n,\delta)$ introduced in \eqref{eq_eX024} and \eqref{eq_eF024}, it is convenient to use the following notation:
\begin{equation}
    e_{F0}^{r,s} = \left. \dfrac{\partial^{\, r+s} e_F\!\left( n(x), \delta \right)}{\partial x^r \, \partial \delta^s}
    \right\rvert_{\substack{x=0\\ \delta=0\\ \vphantom{|}}} 
    \qquad  \qquad \qquad
    e_{F1}^{r,s} = \left. \dfrac{\partial^{\, r+s} e_F\!\left( n(x), \delta \right)}{\partial x^r \, \partial \delta^s}
    \right\rvert_{\substack{x=0\\ \delta=1\\ \vphantom{|}}} 
\end{equation}
where $n(x)=n_0(3x+1)$. A zero value of either $r$ or $s$ indicates no derivative with respect to that variable, e.g.,~$e_{F0}^{0,0}=e_F(n_0,0)$ and~$e_{F1}^{0,0}=e_F(n_0,1)$.

Note that, since we have no $u_1(n)$ contribution, we cannot independently fix the parameters $\chi_1$ to specified values (possibly zero, as done in the metamodel proposed in \citep{Huth2021}), but this is in line with our assumption that their nonzero values arise only from the mass difference~\citep{Haensel_1977}.
\\
\\
\emph{Isoscalar sector} - The parameters $h_0^{(0,1,2)}$ in $u_0$ are related to the isoscalar NMPs $\chi_0$ and $n_0$ by
\begin{equation}
\label{eq_h_coefficients}
\begin{aligned}
    h_0^{(0)} &= E_0 - e_{F0}^{0,0}
    \\
    h_0^{(1)} &= L_0 - e_{F0}^{1,0} + \left(a_0 + b_0 + c_0 \right) \left(E_0 - e_{F0}^{0,0} \right) 
    \\
    h_0^{(2)} &= \frac{1}{2} \left(K_0 - e_{F0}^{2,0} \right) 
    + \left(a_0 + b_0 + c_0 \right) \left( L_0 - e_{F0}^{1,0} \right)
    + \left(a_0 b_0 + b_0 c_0 + c_0 a_0 \right) \left(E_0 - e_{F0}^{0,0} \right).
\end{aligned}
\end{equation}
Despite $L_0=0$ exactly, we keep it explicit in \eqref{eq_h_coefficients} to highlight the similarities with the expressions below for the parameters $h_2^{(0,1,2)}$ that govern the isovector contribution~$u_2$. 
\\
\\
\emph{Isovector sector} - Once the NMPs $n_0$, $\chi_0$ and $\chi_2$ are given, the $h_2^{(0,1,2)}$ parameters are obtained as
\begin{equation}
\label{eq_l_coefficients}
\begin{aligned}
    h_2^{(0)} &= E_2 - \frac{1}{2}  e_{F0}^{0,2} 
    \\
    h_2^{(1)} &= L_2 - \frac{1}{2} e_{F0}^{1,2} 
    + \left( a_2 + b_2 + c_2 \right) \left( E_2 - \frac{1}{2} e_{F0}^{0,2} \right) 
    \\
    h_2^{(2)} &= \frac{1}{2} \left( K_2 -\frac{1}{2}  e_{F0}^{2,2} \right)
    + (a_2 + b_2 + c_2) \left(L_2 - \frac{1}{2}e_{F0}^{1,2} \right) 
    + \left( a_2 b_2 + b_2 c_2 + c_2 a_2 \right) \left( E_2 - \frac{1}{2} e_{F0}^{0,2} \right) .
    \end{aligned}
\end{equation}
Since the mapping in \eqref{eq_h_coefficients} and \eqref{eq_l_coefficients} depends only on the expansion up to $\delta^2$ around $\delta=0$, the presence of $u_4$ in \eqref{eq_eX024} does not alter it.
\\
\\
\emph{Linear terms in $\delta$ at saturation} - Our procedure does not impose a mapping or constraints involving the three coefficients $\chi_1$ in \eqref{eq_snm}, which are given by
\begin{equation}
\label{eq_E1L1K1}
E_1 = \left. \dfrac{\partial e_X}{\partial \delta}
      \right\rvert_{\substack{x=0\\ \delta=0\\ \vphantom{|}}} 
= e_{F0}^{0,1} \, ,
\qquad
L_1 = \left. \dfrac{\partial^2 e_X}{\partial x\partial \delta}
      \right\rvert_{\substack{x=0\\ \delta=0\\ \vphantom{|}}} 
= e_{F0}^{1,1} \, ,
\qquad
K_1 = \left. \dfrac{\partial^3 e_X}{\partial x^2\partial \delta}
      \right\rvert_{\substack{x=0\\ \delta=0\\ \vphantom{|}}} 
= e_{F0}^{2,1} \, .
\end{equation}
As in the original metamodel \citep{MargueronMetaI}, the $\chi_1$ are not forced to vanish, unless one sets $m_n=m_p$ in \eqref{eq_eF024}. 
In practice, the $\chi_1$ coefficients are small but not necessarily zero in our formulation due to the neutron-proton mass difference and the choice of retaining only corrections to the free Fermi gas mixture associated with even powers of $\delta$. This is different from the metamodel of \citet{Huth2021}, where the authors implement a linear correction in the proton fraction (corresponding to a nonzero $u_1(n)$ contribution to $e_X$ in our scheme) to partially correct for the neutron-proton mass difference.
\\
\\
\emph{Mapping for the quartic term} - The quartic contribution $u_4$ in \eqref{eq_eX024} is introduced to control the PNM behavior at saturation independently of the mapping in \eqref{eq_h_coefficients} and \eqref{eq_l_coefficients}, which only fixes the expansion of $e_X(n,\delta)$ around $\delta=0$. 
The two parameters $A$ and $B$ in \eqref{eq_u4} are fixed by requiring that the model reproduces two chosen PNM quantities at saturation, namely the energy per baryon $\tilde{E}$ and its first derivative with respect to $x$ at $\delta=1$, i.e., the parameter $\tilde{L}$ in \eqref{eq_pnm}.
For $A\neq0$, the explicit expressions for $A$ and $B$~are
\begin{equation}
\label{eq_AB}
A = 2\left[ 
\tilde{E}- e_{F1}^{0,0} - E_0 + e_{F0}^{0,0} - E_2   + \frac{1}{2} e_{F0}^{0,2} 
\right], 
\quad \quad
B = 2 - \frac{4}{3A}
\left[ 
\tilde{L} - e_{F1}^{1,0} - L_0 + e_{F0}^{1,0} - L_2 + \frac{1}{2} e_{F0}^{1,2}
\right].
\end{equation}
If $A=0$, the quartic term vanishes and $B$ is irrelevant. Moreover, the mapped value of $B$ must satisfy the condition $B>0$ imposed in \eqref{eq_u4}. The above expressions show that, for $A\neq0$, the quartic term $u_4$ can be calibrated to a chosen PNM reference at saturation without modifying the exact mapping involving the traditional NMPs $\{n_0,E_0,K_0,E_2,L_2,K_2\}$ given in \eqref{eq_h_coefficients} and~\eqref{eq_l_coefficients}.

\section{Priors for $\{a_0,b_0,c_0\}$ and $\{a_2,b_2,c_2\}$}
\label{app_abc}

The parameters $a_i$, $b_i$, and $c_i$ $(i=0,2)$ in \eqref{eq_u02} are sampled using a procedure that removes the degeneracy of $u_i(n)$ under permutation of these three parameters. Directly sampling $a_i$, $b_i$, and $c_i$ uniformly within the cube $[0,3)^3$ would be inefficient, since six permutations of the same triplet produce an identical $u_i(n)$. To avoid this six-fold redundancy, we restrict the sampling to the domain~$D = \{\,0\leq a_i < b_i \leq c_i<3 \,\}$ by defining
\begin{equation}
\label{badessacluniacense}
    \begin{split}
        a_i & = 3\,x_i^{p} ,\\
        b_i & = a_i + (3-a_i)y_i^{p}
              = 3\!\left[1-(1-x_i^{p})(1-y_i^{p})\right] ,\\
        c_i & = b_i + (3-b_i)z_i^{p}
              = 3\!\left[1-(1-x_i^{p})(1-y_i^{p})(1-z_i^{p})\right] ,
    \end{split}
\end{equation}
where $x_i$, $y_i$, and $z_i$ are independent and uniformly distributed in $[0,1)$.
The parameter $p>0$ controls the probability density function $f(a_i,b_i,c_i)$ over $D$, which reads
\begin{equation}
\label{priore_abc}
    f(a_i,b_i,c_i)
    = \frac{\left[a_i(b_i-a_i)(c_i-b_i)\right]^{1/p-1}}
    {3^{1/p}\,p^3\left[(3-a_i)(3-b_i)\right]^{1/p}}\, ,
    \qquad \quad\int_D \! da \, db \, dc\,f(a,b,c) =1\, .
\end{equation}
The sampling of $f(a_i,b_i,c_i)$ is achieved by extracting $x_i$, $y_i$, and $z_i$ independently and uniformly in $[0,1)$ and using~\eqref{badessacluniacense}.

In principle, distinct parameters $p_i$ could be used for the two sets $\{a_0,b_0,c_0\}$ and $\{a_2,b_2,c_2\}$, leading to two different distributions $f_i$. However, this additional freedom is unnecessary for the present work, and we adopt a single value $p=3$. This choice biases the sampling toward smaller values of $\{a_i,b_i,c_i \}$, a natural configuration because the denominators in $u_i(n)$ act mainly as high-density curvature corrections and should remain close to unity near saturation while still enforcing asymptotic causality.  
Conversely, choosing $0<p<1$ would bias the sampling toward $3$, potentially increasing the rejection rate of non-physical models. Therefore, based on trial and error, we adopt $p=3$ as a convenient choice for the prior~\eqref{priore_abc}.

\section{Analytic behavior of $V_i(x)$ around saturation and a low-density alternative}
\label{app_g}

It is convenient to choose the functions $V_i(x)$ in \eqref{eq_u02} in a way that they do not spoil the map between the NMPs $\{ n_0, E_0,K_0, E_2,L_2,K_2 \}$ and $\{h_i^{(0)}, h_i^{(1)},  h_i^{(2)} \}$ described in App.~\ref{app_mapping}. A relatively simple choice that satisfies this requirement is given in~\eqref{eq_V02}, provided that~$g_i>-1$.

The more restrictive condition $g_i \geq 0$ guarantees continuity of the sound speed in a neighborhood of $x=0$, while $g_i>0$ also guarantees continuity of its first derivative. To see this, recall that pressure and chemical potentials depend only on the first derivatives of the energy per baryon with respect to $x$, whereas the sound speed also depends on the second derivative. Requiring continuity of the first derivative of the sound speed further forces continuity of the third derivative of $V_i(x)$.

For a positive integer $k$ and a real $q>0$, we have $\lim_{x\to 0^{\pm}} d^k(\pm x)^q/dx^k = 0$ when $k<q$, and $\lim_{x\to 0^{\pm}} d^k(\pm x)^q/dx^k = (\pm1)^k k!$ when $q=k$. Thus, the expansion of $V_i$ near saturation reads
\begin{equation}
\label{eq_V02x}
    V_i(x) = s_i |3x|^{g_i}
    \left( \frac{|3x|^3}{1 + w_i} + O(x^4) \right) \, .
\end{equation}
For $g_i>0$, this implies (the prime denotes differentiation with respect to $x$)
\begin{equation}
\label{eq_V02x_lim}
    \lim_{x\rightarrow 0^-} V_i'''(x)
    = \lim_{x\rightarrow 0^+} V_i'''(x)
    = 0 \, .
\end{equation}
For $g_i=0$, $V_i^{\prime\prime\prime}(x)$ has a finite jump at saturation, so the sound speed remains continuous but its first derivative need not be. For $0 < g_i < 1$, the fourth derivative diverges as $x\to0^{\pm}$; for $g_i = 1$ it remains finite, and for $g_i > 1$ it vanishes. Thus, choosing a larger value of $g_i$ increases the smoothness of the energy per baryon and of all thermodynamic quantities. For instance, requiring $g_i > j$ (for integer $j\geq0$) guarantees that both the pressure and the chemical potentials are continuous up to the $(2+j)$-th derivative in $x$, and that the sound speed is continuous up to at least the $(1+j)$-th derivative.

Finally, one may replace $V_i(x)$ by $V_i(x)\theta(-x)$ in \eqref{eq_u02}, where $\theta$ is the usual unit step function. For $x<0$, the local expansion in \eqref{eq_V02x} of $V_i(x)$ is unchanged, whereas $V_i(x)=0$ for $x>0$. Consequently, the NMP mapping and the continuity properties of the sound speed discussed above remain unchanged: $g_i\geq0$ guarantees continuity of the sound speed, while $g_i>0$ also guarantees continuity of its first derivative. For $g_i=1$, the fourth derivative remains finite but generally develops a jump at~$x=0$.

This defines an alternative parametrization in which the parameters $w_i$ and $g_i$ are completely decoupled from the suprasaturation regime $x>0$. An exploratory analysis shows that neither the quality of the fits in Sec.~\ref{sec_fits} nor the Bayesian posterior distributions reported in Sec.~\ref{sec_result} change appreciably. Although this provides only a limited robustness check, since the modification is minimal, it is nevertheless informative: it shows that ad hoc low-density corrections of the kind proposed in \cite{burrello2025crust}, which are completely decoupled from the suprasaturation behavior, can be implemented in future studies.


\bibliography{Biblio}

\label{lastpage}
\end{document}